**Operant Conditioning in Indian Free-Ranging Dogs: Effects of Positive and Threatening Cues on Sociability**

Srijaya Nandi[a], Dipanjan Roy[a], Abantika Adak[b], Rohan Sarkar[a], Anindita Bhadra[a*]

[a]Department of Biological Sciences, Indian Institute of Science Education and Research, Kolkata, Mohanpur, Nadia, West Bengal, India

[b]Department of Biotechnology, Maulana Abul Kalam Azad University of Technology, Haringhata, Nadia, West Bengal, India

Corresponding Author:

Anindita Bhadra[a*]

abhadra@gmail.com

Email IDs:

sn21rs025@iiserkol.ac.in, dipanjanroy0520@gmail.com, abantikaadak2003@gmail.com, rohansarkar2191@gmail.com, abhadra@iiserkol.ac.in

## Abstract

Sociability toward humans is a key adaptive trait in free-ranging dogs, enabling them to access resources while navigating risks associated with human interactions. In this study, we investigated whether operant conditioning shapes sociability in Indian free-ranging dogs and whether learned responses generalize to unfamiliar individuals. We experimentally exposed 58 dog groups to either positive or a threatening cue over five consecutive days and assessed their behaviour using approach proportion, approach latency, and demeanor across repeated interactions with a familiar experimenter, followed by a test with an unfamiliar individual.

Using Bayesian generalized linear mixed models, we found that cue type and repeated exposure significantly influenced sociability. Dogs exposed to a positive cue showed increased approach behaviour and reduced approach latency over time, along with increased affiliative demeanor. In contrast, dogs exposed to threatening cues exhibited reduced approach behaviour, increased approach latency, and a shift toward neutral and less affiliative responses across days.

Importantly, positive cues partially generalized across individuals, as dogs showed increased approach toward an unfamiliar experimenter, although this was accompanied by hesitation to approach. In contrast, threatening cues did not generalize in the same way: dogs did not reduce their approach toward unfamiliar individuals but displayed increased approach latency, indicating heightened caution.

Overall, our findings demonstrate that operant conditioning plays a crucial role in shaping dog-human interactions, with asymmetric generalization of positive and threatening experiences. This behavioural flexibility likely reflects an adaptive strategy that allows free-ranging dogs to balance resource acquisition with risk avoidance in human-dominated environments. These insights have important implications for improving human-dog coexistence and informing management strategies for free-ranging dog populations.

## Introduction

Sociability is a core personality trait in animals (Réale et al., 2007), defined as the tendency of an individual to associate with others in contexts not directly related to reproduction or aggression (Gartland et al., 2022). Such associations can occur both within and across species and may serve a range of adaptive functions, including reducing predation risk, facilitating ectoparasite removal, improving foraging success, and conserving energy (Krause & Ruxton, 2002). For animals coexisting with humans, as in the case of urbanized and domesticated animals, sociability towards people can be particularly beneficial, as it often improves access to resources. However, approaching unfamiliar humans carries inherent risks, and poor decision-making in this regard, can lead to severe negative outcomes.

One mechanism that may shape sociability in these contexts is operant conditioning, a form of learning in which the future likelihood of a behaviour depends on the immediate consequence it leads to (Sproatt & Navab, 2013). Behaviours followed by reinforcement tend to increase in frequency, whereas those followed by punishment typically decrease (Powell et al., 2022). Operant conditioning has been documented across a broad range of taxa, from invertebrates (Brembs, 2003) to vertebrates (Staddon and Cerutti, 2003). It plays a major role in the development of behavioural plasticity among animals, along with other factors such as evolution and cultural transmission (Mery and Burns, 2010).

Learning by operant conditioning has been studied in several domesticated species such as pet dogs (Jezierski, 2010), cats (Delgado and Reevy, 2024), goats (Wenzel et al., 1964), cows (Vaughan et al., 2014) and horses (Ninomiya et al., 2007). Free-ranging dogs represent a particularly interesting population for studying the phenomenon of operant conditioning. Unlike companion dogs, they are not directly supervised by humans, and their daily movements and activities occur independently of human control (Boitani et al., 2017). They have coexisted alongside humans for centuries across all possible human habitats (Berman & Dunbar, 1983) and account for roughly 80% of the global dog population (Boitani and Ciucci, 1995; Hughes and Macdonald, 2013). They are largely dependent on human-generated waste for their sustenance and also tend to give birth near human habitations (Sen Majumder et al., 2016). Their close proximity to people exposes them to a mixture of positive and negative human interactions, which, in turn, can strongly influence their behavioural responses (Bhattacharjee et al., 2021). Previous research has shown that Indian free-ranging dogs increase their sociability toward humans over repeated interactions (Bhattacharjee et al., 2017; Nandi et al., 2024), and adult free-ranging dogs respond differently to positive and negative reinforcement in pointing gesture following tasks. However, such studies focused on dogs that initially approached an unfamiliar experimenter, leaving open the question of whether sociability can also be modified in non-approachable individuals.

In the present study, we investigated whether operant conditioning contributes to the sociability of Indian free-ranging dogs towards unfamiliar humans. Specifically, we examined two contingencies that they commonly encounter in their daily environment: receiving food (positive cue) and exposure to an aversive stimulus such as a stick (threatening cue). We further aimed to assess whether the rate at which these forms of conditioning develop differs, thereby providing insight into the learning dynamics underlying dog-human interactions. This comparison is critical, as organisms often exhibit a 'negativity bias,' where learning to avoid a threat occurs more rapidly and robustly than learning to approach a reward, given the higher potential cost of a negative outcome (Baumeister et al., 2001). Finally, we asked whether changes in sociability generalize across individuals, or whether they remain specific to the person involved in repeated interactions. Our study, therefore, addressed four key questions: (a) does a threatening cue reduce sociability, (b) does a positive cue enhance sociability, (c) does the level of learning differ between the positive and threatening cues, and d) do these effects extend to unfamiliar humans or remain tied to the familiar experimenter?

Based on the principles of operant conditioning, we formulated three primary hypotheses. First, we predicted that dogs in the threatening cue group would show a significant decrease in sociability toward the familiar experimenter, as the aversive stimulus would discourage approach behaviour. Second, we hypothesized that dogs in the positive cue group would display a significant increase in sociability, as they learn to associate the experimenter with a food reward. Third, we hypothesized that the rate of learning would be faster for the threatening cue groups compared to positive cue, based on the principle of "negativity bias". Finally, concerning generalization, we predicted that these learned responses would be specific to the familiar experimenter. We did not expect a significant change in sociability toward the unfamiliar experimenter, suggesting that dogs would discriminate between the person associated with the consequence and a neutral stranger.

To test these predictions, we assigned free-ranging dog groups to one of two experimental groups: a food reward (positive cue) and a stick display (threatening cue), and systematically measured changes in their sociability over repeated trials with both a familiar and an unfamiliar experimenter.

**Methodology**

## Ethics Statement

The study design did not violate the Animal Ethics regulations of the Government of India (Prevention of Cruelty to Animals Act 1960, Amendment 1982). The experimental protocol was approved by the IISER Kolkata Animal Ethics Committee, as part of a larger project sanctioned by the SERB (EMR/2016/000595).

## Subjects

We identified 58 free-ranging dog groups for the study. Groups were defined as three or more dogs located within approximately five meters of each other on the first day of the experiment. Inclusion criteria required a minimum of three visibly healthy adults per group. A male was classified as an adult if his testes were descended, and a female was classified as an adult if her nipples were darkened. In total, 29 groups were randomly assigned to receive a positive cue and 29 groups to receive a threatening cue. All dogs were photographed for later identification, and group locations were recorded using a GPS-enabled mobile phone.

## Study Areas

Experiments were conducted in four urban and semi-urban sites in West Bengal, India: Gayeshpur (22°57′29″N, 88°29′33″E), Anandanagar (22°58′44″N, 88°29′15″E), Kataganj (22°56′58″N, 88°28′14″E), and Kanchrapara ( 22°56'35.13"N,  88°26'10.01"E). A map showing the study areas is provided in Figure S1.

## Experimenters involved

Three experimenters, all unfamiliar to the focal dog groups, were involved. Experimenter 1 (E1, female) conducted the experiments from Days 1-6. Experimenter 2 (E2, female) was involved only on Day 6 and acted as the unfamiliar person. Experimenter 3 (E3, male) was responsible for video recording throughout the study.

## Experimental Procedure

### Approachability Test (Days 1-5)

On Day 1, E1 initiated an approachability test by standing ~1.5 m from the nearest dog in a group. She produced a standardized positive vocalization, “ae-ae-ae” (Bhattacharjee et al., 2017) for 5 s to attract the attention of the dogs, and then maintained eye contact with the dogs for 25 s (Video S1). Each group then received either a positive or a threatening cue, depending on its assignment. The same protocol was repeated from Day 2 to Day 5. After the approachability test, the dog groups were exposed to either the positive or the threatening cue.

**Positive cue:** Following the approachability test, E1 placed Parle-G (glucose) biscuits near the dogs that approached (Video S2). The number of biscuits matched the number of dogs that had approached. She then threw biscuits toward the non-approaching dogs from a distance, ensuring that the dogs had to approach the biscuits to interact with them. Dogs were allowed up to 30s to respond to the biscuits, which included behaviours like sniffing, licking, or eating.

**Threatening cue:** Following the approachability test, E1 displayed a 45 cm wooden stick for 30 s to the dogs that had approached (Video S3). Dogs that did not approach could also observe the stick from a distance.

The approachability test, followed by the administration of either the positive and threatening cues was done from Day 1 to Day 5.

**Approachability Test by an Unfamiliar Person (Day 6)**

On Day 6, E2 (unfamiliar to the dog groups) conducted the approachability test using the same procedure as E1. Immediately afterwards, she left the site and remained hidden. After a 5-minute interval, E1 repeated the test with the same group. This design allowed us to examine whether changes in sociability can be generalised to unfamiliar individuals or remain specific to the familiar experimenter.

The approachability tests conducted on Day 6 by E2 and E1 were labelled 6A and 6B respectively. The experimental protocol is illustrated in Figure 1.

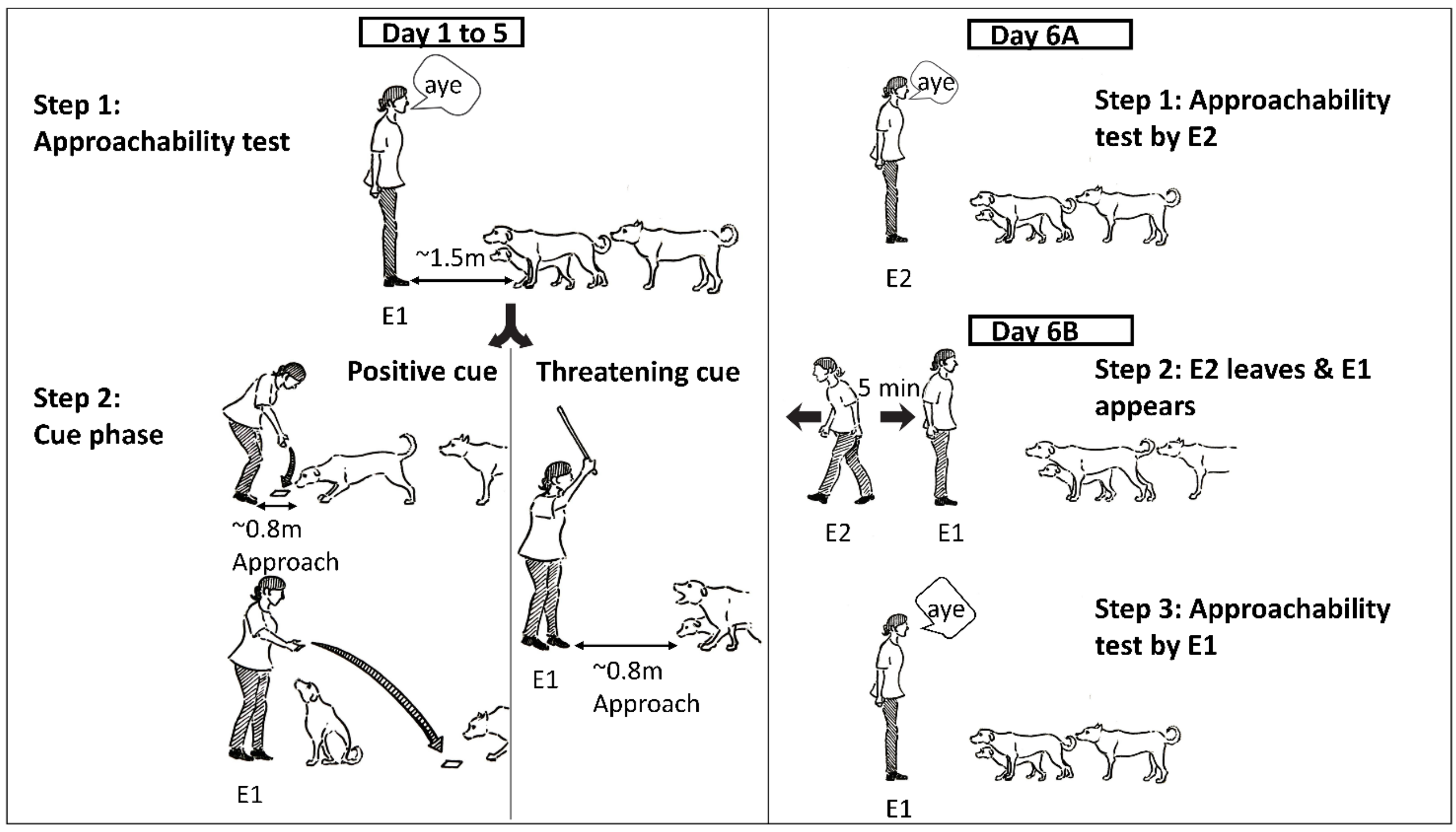

**Figure 1.** Schematic representation of the experimental protocol. Illustration by Arpan Bhattacharyya.

**Data decoding**

The following behavioural parameters were coded from video recordings:

1) **Proportion of approach (approachability test):** An approach was considered when a dog came within one dog body length (~0.8 m) of the experimenter. The proportion of approach was calculated as the number of dogs that approached divided by the total number of dogs present at the start of the test.

2) **Approach latency (approachability test):** Approach latency was defined as the time (in seconds) taken by a dog to approach within one dog body length of the experimenter from the moment the experimenter first called the dogs. Dogs that did not approach within the allotted time were treated as censored, and their approach latency was recorded as the maximum allowed duration to approach (30s).

3) **Demeanor (approachability and test phases):** Demeanor was classified following Bhattacharjee et al. (2021). Only the initial demeanor displayed during each phase was considered.

All videos were coded by D.R. A.A. independently coded 20% of the videos selected at random to assess inter-rater reliability. Agreement between observers was high for both approach behaviour (Cohen's $\kappa = 0.87$) and demeanor classification (Cohen's $\kappa = 0.85$).

**Statistical analysis**

All analyses were conducted in a Bayesian framework using generalized linear mixed-effects models (GLMMs) in the brms package (Bürkner, 2018) in R version 4.2.3 (R Core Team, 2023). We used weakly informative priors specifying a normal distribution (mean = 0, SD = 1) on the fixed effects and an exponential distribution (SD = 1) on the random effects for all models (Model specific shape parameter priors have been mentioned in relevant sections). Thus, we favoured no specific direction for the effects while keeping their magnitude within reasonable values. We used 95% HPD interval to assess the variability in parameter estimates and to interpret our findings. We reported differences between levels using contrasts (for multinomial models) and pairwise contrasts using emmeans (for zero-one inflated beta and lognormal survival models). The contrasts were reported in response scale (i.e., probability for logistic regression). HPD interval that did not overlap zero indicated a difference between parameters. For models, we ran 5,000 (8,000, for complex models) iterations of the Markov Chain and discarded the first 1,000 (or 2000) as a warm-up.

Pairwise comparisons were restricted to contrasts of primary interest: Day 1 versus all subsequent days (Days 2-5, 6A, and 6B), and a comparison between Day 6A and Day 6B.

The four models were as follows:

**Model 1: Zero-one inflated beta regression model examining factors affecting approach proportion during the approachability test**

The proportion of dogs that approached during the approachability test (approach_proportion) was analysed using a zero-one inflated beta regression model. The fixed effects included day (seven levels: Day 1, Day 2, Day 3, Day 4, Day 5, Day 6A, and Day 6B), cue (two levels: positive and threatening), and their interaction (day × cue). A random intercept for group identity (group_id) was included to account for repeated observations within groups. Pairwise comparisons were reported as median difference in the expected proportion of approach with 95% HPD interval.

The model structure was:

$$\text{approach_proportion} \sim \text{day} \times \text{cue} + (1 \mid \text{group_id})$$

**Model 2: Lognormal survival model examining factors affecting approach latency during the approachability test)**

The latency to approach (time_pv) was analysed using a lognormal survival model, accounting for right-censoring via the "cens_brms" indicator. The fixed effects included day, cue, sex, and the interaction between day and cue.  To account for the hierarchical structure of the data, random intercepts were included for group identity (group_id) and individual dogs nested within groups (group_dog). Pairwise comparisons were reported as median difference in expected approach latency (in seconds) with 95% HPD interval.

The model structure was:

$$\text{time_pv} \mid \text{cens(cens_brms)} \sim \text{day} \times \text{cue} + \text{sex} + (1 \mid \text{group_id/group_dog})$$

**Model 3: Multinomial regression model examining factors affecting demeanor during the approachability test and**

**Model 4: Multinomial regression model examining factors affecting demeanor during the test phase**

Demeanor, a categorical response variable, was analysed using a multinomial regression model. The fixed effects included cue, day, sex, and the interaction between cue and day.

Model 3 included 7 levels for day (Day 1, Day 2, Day 3, Day 4, Day 5, Day 6A and Day 6B), while model 4 included 5 levels for day (Day 1, Day 2, Day 3, Day 4 and Day 5).

To account for the hierarchical structure of the data, random intercepts were included for group identity (group_id) and individual dogs nested within groups (group_dog). Estimated marginal contrasts of predicted probabilities were reported as median difference in predicted probability between conditions, with 95% HPD interval used to quantify uncertainty.

The model structure was:

$$\text{demeanor} \sim \text{cue} \times \text{day} + \text{sex} + (1 \mid \text{group_id/group_dog})$$

## Results

### 1) What factors affect the proportion of approach during the approachability test?

The regression model revealed a strong interaction between cue type and test day in shaping approach behaviour (Table 1).

**Table 1.** Posterior parameter estimates (mean, SE, and 95% credible intervals) from the zero-one-inflated beta model assessing the effects of day, cue type, and their interaction on approach proportion, including model diagnostics ($\hat{R}$, Bulk ESS, Tail ESS). Parameters marked with an asterisk (*) indicate statistically credible effects where the 95% CI completely excludes zero.

| Parameter | Estimate | Est.Error | l-95% CI | u-95% CI | Rhat | Bulk_ESS | Tail_ESS |
|---|---|---|---|---|---|---|---|
| Intercept | -0.29 | 0.16 | -0.61 | 0.03 | 1.00 | 4914 | 8231 |
| dayDay2 | 0.21 | 0.24 | -0.25 | 0.68 | 1.00 | 7027 | 9990 |
| dayDay3 | 0.14 | 0.19 | -0.23 | 0.51 | 1.00 | 6293 | 9717 |
| dayDay4 | 0.15 | 0.21 | -0.27 | 0.55 | 1.00 | 6704 | 8994 |
| dayDay5 | 0.23 | 0.22 | -0.20 | 0.65 | 1.00 | 6830 | 10202 |
| **dayDay6A *** | **0.48** | **0.22** | **0.04** | **0.91** | **1.00** | **6848** | **10440** |
| dayDay6B | 0.31 | 0.21 | -0.11 | 0.72 | 1.00 | 6908 | 9999 |
| **cueTHREATENING *** | **0.42** | **0.20** | **0.02** | **0.82** | **1.00** | **5121** | **8355** |
| dayDay2:cueTHREATENING | -0.48 | 0.31 | -1.08 | 0.13 | 1.00 | 7487 | 9955 |
| dayDay3:cueTHREATENING | -0.51 | 0.28 | -1.07 | 0.04 | 1.00 | 7836 | 10611 |
| dayDay4:cueTHREATENING | -0.32 | 0.27 | -0.85 | 0.20 | 1.00 | 7018 | 9925 |
| **dayDay5:cueTHREATENING *** | **-0.66** | **0.28** | **-1.21** | **-0.09** | **1.00** | **7133** | **10151** |
| **dayDay6A:cueTHREATENING *** | **-0.65** | **0.30** | **-1.23** | **-0.06** | **1.00** | **7157** | **11058** |

| dayDay6B:cueTHREATENING * | -0.86 | 0.29 | -1.41 | -0.29 | 1.00 | 7177 | 10170 |
|---|---|---|---|---|---|---|---|

To interpret this interaction, we examined estimated marginal means, which represent model-predicted effects averaged over other predictors.

**a) What is the difference in the approach proportion within day between the positive and threatening cues?**

On Day 1, approach was lower for groups exposed to the positive cue than those exposed to the threatening cue (median estimate = -0.10; 95% HPD: -0.20, -0.005; Figure 2a).

**b) What is the difference in approach proportion across days within the positive cue?**

Approach proportion was higher for the positive cue groups, most notably on Day 6A compared to Day 1 (median estimate = -0.12; 95% HPD: -0.23, -0.02; Figure 2b).

**c) What is the difference in approach proportion across days within the threatening cue?**

Approach proportion on Day 1 was higher than on Day 5 (median estimate = 0.10; 95% HPD: 0.01, 0.20) and Day 6B (median estimate = 0.14; 95% HPD: 0.04, 0.24; Figure 2c). All the pairwise comparisons are presented in Table S1.

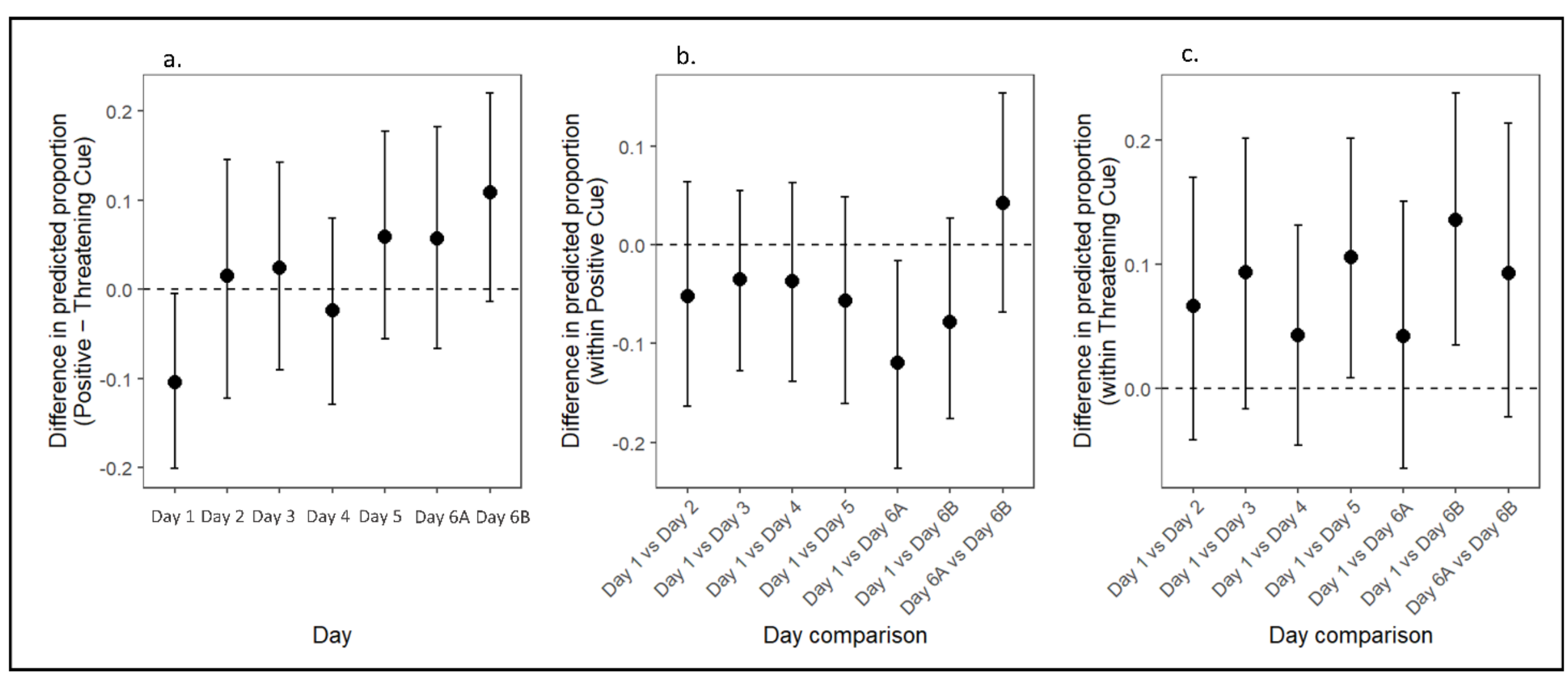

**Figure 2.** Difference in predicted proportion for a) Between-cue contrasts (Threatening - Positive Cue) across days, b) Within-positive cue contrasts between selected day comparisons (relative to Day 1 and between Days 6A and 6B), c) Within-threatening cue contrasts showing analogous day-to-day changes under the threatening cue condition. Points represent posterior medians and error bars indicate 95% HPD. The dashed horizontal line at zero indicates no difference.

**2) What factors affect the approach latency during the approachability test?**

The lognormal mixed-effects survival model revealed clear effects of test day, cue type, their interaction and sex on approach latency (Table 2).

**Table 2.** Posterior parameter estimates (mean, SE, and 95% credible intervals) from the Bayesian regression model assessing the effects of day, cue type, sex, and their interaction on the response variable, including model diagnostics ($\hat{R}$, Bulk ESS, Tail ESS). Parameters marked with an asterisk (*) indicate statistically credible effects where the 95% CI completely excludes zero.

| Parameter | Estimate | Est.Error | l-95% CI | u-95% CI | Rhat | Bulk_ESS | Tail_ESS |
|---|---|---|---|---|---|---|---|
| **Intercept *** | **4.50** | **0.27** | **3.98** | **5.03** | **1.00** | **6421** | **9291** |
| DAYDay2 | -0.10 | 0.26 | -0.61 | 0.42 | 1.00 | 8252 | 10591 |
| DAYDay3 | -0.27 | 0.26 | -0.76 | 0.24 | 1.00 | 7879 | 10049 |
| **DAYDay4 *** | **-0.83** | **0.24** | **-1.30** | **-0.35** | **1.00** | **7642** | **10853** |
| **DAYDay5 *** | **-0.69** | **0.25** | **-1.19** | **-0.21** | **1.00** | **7865** | **10866** |
| DAYDay6A | -0.28 | 0.26 | -0.79 | 0.22 | 1.00 | 7884 | 10577 |
| **DAYDay6B *** | **-0.51** | **0.25** | **-1.01** | **-0.01** | **1.00** | **7843** | **11536** |
| **CUETHREATENING *** | **-0.78** | **0.33** | **-1.43** | **-0.14** | **1.00** | **6134** | **9017** |
| **SEXM *** | **-0.49** | **0.19** | **-0.86** | **-0.13** | **1.00** | **9466** | **11442** |
| **DAYDay2:CUETHREATENING *** | **1.05** | **0.36** | **0.35** | **1.76** | **1.00** | **9443** | **11124** |
| DAYDay3:CUETHREATENING * | 1.45 | 0.36 | 0.74 | 2.15 | 1.00 | 9262 | 11499 |
| DAYDay4:CUETHREATENING * | 1.58 | 0.34 | 0.89 | 2.24 | 1.00 | 8590 | 10869 |
| DAYDay5:CUETHREATENING * | 1.87 | 0.36 | 1.16 | 2.57 | 1.00 | 9644 | 11369 |
| DAYDay6A:CUETHREATENING * | 0.92 | 0.35 | 0.24 | 1.61 | 1.00 | 8979 | 11036 |
| DAYDay6B:CUETHREATENING * | 1.76 | 0.36 | 1.05 | 2.47 | 1.00 | 9574 | 11652 |

Estimated marginal means further clarified day and cue-specific patterns.

**a) What is the difference in approach latency within day between the positive and threatening cues?**

Day-specific contrasts between cue types showed that on Day 1, approach latency was longer for the positive cue groups than the threatening cue ones (median estimate = 0.78; 95% HPD: 0.14, 1.43). From Day 4 onward, this pattern reversed, with approach latency credibly longer under the threatening cue on Day 4 (median estimate = -0.79; 95% HPD: -1.52, -0.05), Day 5 (median estimate = -1.09; 95% HPD: -1.87, -0.34), and Day 6B (median estimate = -0.97; 95% HPD: -1.74, -0.19; Figure 3a).

**b) What is the difference in approach latency across days within the positive cue?**

Approach latencies on Day 1 were higher than those on Day 4 (median estimate = 0.82; 95% HPD: 0.35, 1.30), Day 5 (median estimate = 0.69; 95% HPD: 0.22, 1.20), and Day 6B (median estimate = 0.50; 95% HPD: 0.005, 1.00; Figure 3b).

**c) What is the difference in approach latency across days within the threatening cue?**

Approach latency was significantly lower on Day 1 compared to subsequent days, including Day 2 (median estimate = -0.94; 95% HPD: -1.48, -0.39), Day 3 (median estimate = -1.18; 95% HPD: -1.74, -0.62), Day 4 (median estimate = -0.75; 95% HPD: -1.28, -0.21), Day 5 (median estimate = -1.18; 95% HPD: -1.76, -0.61), Day 6A (median estimate = -0.64; 95% HPD: -1.15, -0.10), and Day 6B (median estimate = -1.25; 95% HPD: -1.82, -0.69), indicating faster initial approach followed by longer latencies on later days (Figure 3c). All pairwise comparisons are presented in Table S2.

**d) What is the difference in approach latency between male and female dogs?**

Female dogs showed higher approach latency compared to male dogs (median estimate = 0.49; 95% HPD: 0.11, 0.84).

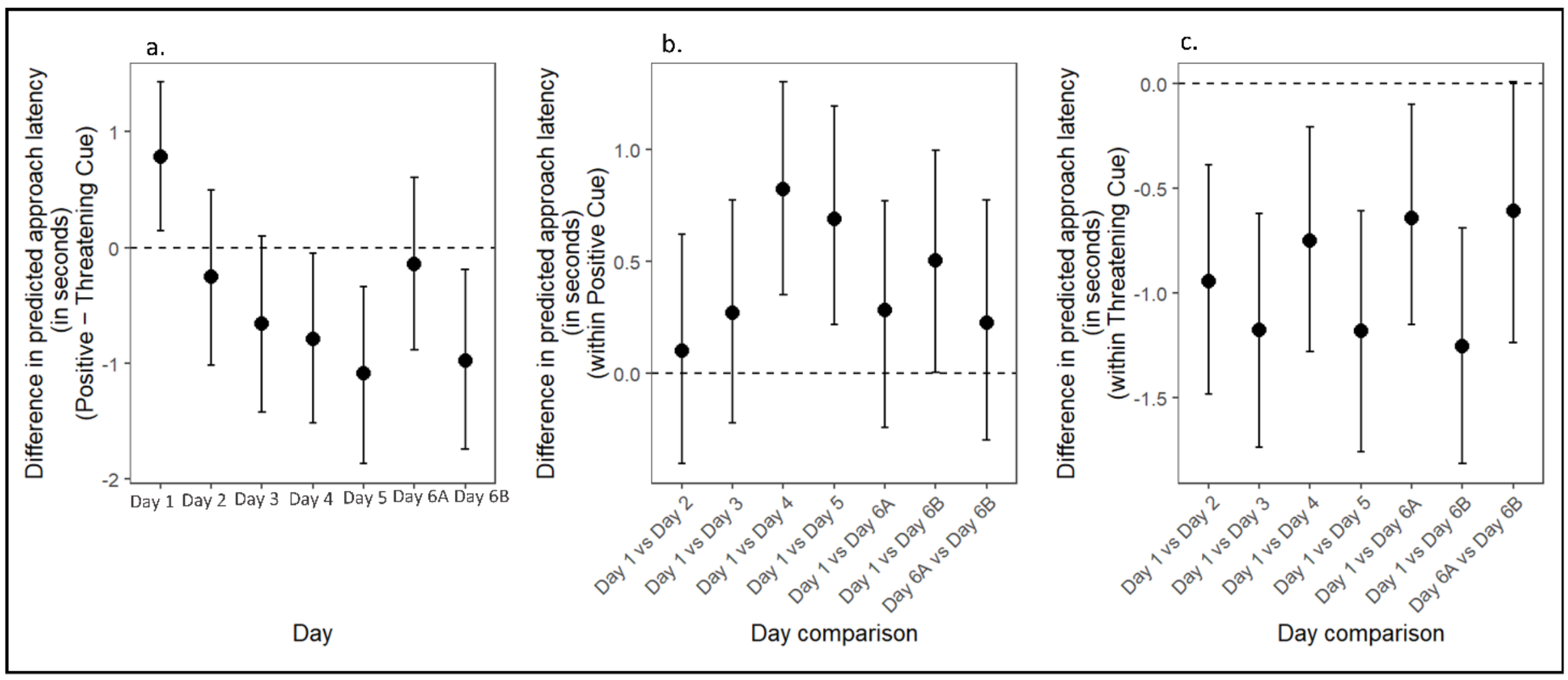

**Figure 3.** Difference in predicted approach latency (in seconds) for a) Between-cue contrasts (Threatening - Positive Cue) across days, b) Within-positive cue contrasts between selected day comparisons (relative to Day 1 and between Days 6A and 6B), c) Within-threatening cue contrasts showing analogous day-to-day changes. Points represent posterior medians and error bars indicate 95% HPD. The dashed horizontal line at zero indicates no difference.

**3) What factors affect demeanor during the approachability test?**

The model revealed significant effects of day and cue on dogs' demeanor during the approachability test (Table 3).

**Table 3:** Posterior parameter estimates (mean, SE, and 95% credible intervals) from the Bayesian multinomial regression model assessing the effects of day, cue type, sex, and their interaction on demeanor categories (aggressive, anxious and neutral, affiliative) during the approachability test, including model diagnostics ($\hat{R}$, Bulk ESS, Tail ESS). "AG", "AN" and "N" represent aggressive, anxious and neutral demeanors compared

against the reference demeanor “AF” or affiliative. Parameters marked with an asterisk (*) indicate statistically credible effects where the 95% CI completely excludes zero.

| Parameter | Estimate | Est.Error | l-95% CI | u-95% CI | Rhat | Bulk_ESS | Tail_ESS |
|---|---|---|---|---|---|---|---|
| **muAG_Intercept *** | **-2.34** | **0.48** | **-3.31** | **-1.42** | **1.00** | **8307** | **10574** |
| **muAN_Intercept *** | **-4.15** | **0.57** | **-5.34** | **-3.09** | **1.00** | **11343** | **9310** |
| **muN_Intercept *** | **0.98** | **0.30** | **0.39** | **1.58** | **1.00** | **6684** | **8828** |
| muAG_CUETHREATENING | 0.49 | 0.51 | -0.51 | 1.47 | 1.00 | 9599 | 11382 |
| muAG_DAYDay2 | -0.33 | 0.48 | -1.28 | 0.59 | 1.00 | 12743 | 12723 |
| muAG_DAYDay3 | -0.84 | 0.53 | -1.92 | 0.16 | 1.00 | 14873 | 11773 |
| muAG_DAYDay4 | -0.90 | 0.50 | -1.90 | 0.04 | 1.00 | 13496 | 12146 |
| **muAG_DAYDay5 *** | **-1.59** | **0.58** | **-2.74** | **-0.49** | **1.00** | **16283** | **12469** |
| muAG_DAYDay6A | -0.89 | 0.48 | -1.86 | 0.04 | 1.00 | 13518 | 11650 |
| **muAG_DAYDay6B *** | **-1.16** | **0.53** | **-2.23** | **-0.17** | **1.00** | **16167** | **12714** |
| muAG_SEXM | 0.13 | 0.34 | -0.53 | 0.80 | 1.00 | 14892 | 12607 |
| muAG_CUETHREATENING:DAYDay2 | 0.35 | 0.58 | -0.79 | 1.50 | 1.00 | 13825 | 12934 |
| muAG_CUETHREATENING:DAYDay3 | -0.08 | 0.63 | -1.31 | 1.19 | 1.00 | 15889 | 11503 |
| muAG_CUETHREATENING:DAYDay4 | -0.38 | 0.63 | -1.63 | 0.85 | 1.00 | 13942 | 11975 |
| muAG_CUETHREATENING:DAYDay5 | 0.26 | 0.70 | -1.13 | 1.62 | 1.00 | 17089 | 12423 |
| muAG_CUETHREATENING:DAYDay6A | 0.35 | 0.59 | -0.81 | 1.52 | 1.00 | 13644 | 12348 |
| muAG_CUETHREATENING:DAYDay6B | -0.19 | 0.71 | -1.60 | 1.17 | 1.00 | 16451 | 12743 |
| **muAN_CUETHREATENING *** | **1.10** | **0.54** | **0.04** | **2.15** | **1.00** | **13277** | **12115** |
| muAN_DAYDay2 | 0.28 | 0.68 | -1.10 | 1.59 | 1.00 | 14647 | 11990 |
| muAN_DAYDay3 | 0.70 | 0.63 | -0.53 | 1.90 | 1.00 | 13387 | 11674 |
| muAN_DAYDay4 | -0.06 | 0.67 | -1.39 | 1.23 | 1.00 | 15136 | 11870 |
| muAN_DAYDay5 | -0.36 | 0.72 | -1.82 | 1.02 | 1.00 | 15931 | 11389 |
| muAN_DAYDay6A | -0.54 | 0.71 | -1.96 | 0.78 | 1.00 | 16649 | 11829 |
| muAN_DAYDay6B | 0.43 | 0.64 | -0.87 | 1.65 | 1.00 | 14453 | 11747 |
| muAN_SEXM | -0.25 | 0.39 | -1.01 | 0.53 | 1.00 | 17965 | 12183 |
| muAN_CUETHREATENING:DAYDay2 | -0.03 | 0.76 | -1.51 | 1.46 | 1.00 | 15485 | 12313 |
| muAN_CUETHREATENING:DAYDay3 | 0.89 | 0.67 | -0.38 | 2.21 | 1.00 | 13302 | 11784 |
| muAN_CUETHREATENING:DAYDay4 | 0.61 | 0.72 | -0.78 | 2.04 | 1.00 | 15318 | 12052 |
| muAN_CUETHREATENING:DAYDay5 | 0.05 | 0.81 | -1.56 | 1.63 | 1.00 | 17332 | 12477 |
| muAN_CUETHREATENING:DAYDay6A | -0.09 | 0.79 | -1.66 | 1.45 | 1.00 | 18615 | 12758 |
| muAN_CUETHREATENING:DAYDay6B | 0.59 | 0.71 | -0.78 | 1.96 | 1.00 | 15000 | 12344 |
| muN_CUETHREATENING | -0.25 | 0.37 | -0.98 | 0.46 | 1.00 | 6926 | 9082 |
| muN_DAYDay2 | 0.25 | 0.30 | -0.33 | 0.85 | 1.00 | 9437 | 10039 |
| muN_DAYDay3 | 0.28 | 0.29 | -0.29 | 0.87 | 1.00 | 9178 | 11456 |
| muN_DAYDay4 | -0.30 | 0.29 | -0.85 | 0.27 | 1.00 | 9111 | 11383 |
| muN_DAYDay5 | -0.14 | 0.29 | -0.70 | 0.43 | 1.00 | 9175 | 10421 |
| **muN_DAYDay6A *** | **-0.87** | **0.29** | **-1.45** | **-0.31** | **1.00** | **9463** | **11783** |
| muN_DAYDay6B | -0.36 | 0.29 | -0.94 | 0.21 | 1.00 | 9296 | 11566 |
| **muN_SEXM *** | **-0.57** | **0.23** | **-1.02** | **-0.14** | **1.00** | **9095** | **10692** |
| muN_CUETHREATENING:DAYDay2 | 0.28 | 0.41 | -0.54 | 1.09 | 1.00 | 10219 | 11561 |
| muN_CUETHREATENING:DAYDay3 | -0.08 | 0.41 | -0.87 | 0.71 | 1.00 | 10079 | 11283 |
| muN_CUETHREATENING:DAYDay4 | 0.41 | 0.40 | -0.36 | 1.19 | 1.00 | 10058 | 11651 |
| **muN_CUETHREATENING:DAYDay5 *** | **1.14** | **0.41** | **0.32** | **1.95** | **1.00** | **10265** | **11486** |
| **muN_CUETHREATENING:DAYDay6A *** | **1.12** | **0.40** | **0.34** | **1.92** | **1.00** | **9880** | **11151** |
| **muN_CUETHREATENING:DAYDay6B *** | **1.41** | **0.41** | **0.62** | **2.22** | **1.00** | **10118** | **10954** |

The results of pairwise contrasts are described below:

**a) What is the difference in the approachability test demeanor within day between the positive and threatening cues?**

Estimated marginal contrasts of predicted probabilities indicated differences between cue types across most demeanour categories and test days.

Anxious demeanor was higher under the threatening cue on Day 3 (median estimate = 0.06, 95% HPD: 0.01, 0.12) and Day 4 (median estimate = 0.02, 95% HPD: 0.00005, 0.06). Affiliative demeanor was less likely under the threatening cue on Day 5 (median estimate = -0.17, 95% HPD: -0.31, -0.004), Day 6A (median estimate = -0.21, 95% HPD: -0.39, -0.03), and Day 6B (median estimate = -0.22, 95% HPD: -0.40, -0.08). In contrast, neutral demeanor was more likely under the threatening cue on Day 6B (median estimate = 0.21, 95% HPD: 0.06, 0.41; Figure 4a). Overall, these results suggest that while early responses to cue type were largely similar, dogs increasingly showed reduced affiliative demeanor and increased neutral demeanor under the threatening cue with repeated exposure. The results of all pairwise contrasts are presented in Table S3.

**b) What is the difference in approachability test demeanor across days within the positive cue?**

Affiliative demeanor increased over time, with higher probabilities observed on Day 6A relative to Day 1 (median estimate = 0.21, 95% HPD: 0.09, 0.34). Aggressive demeanor showed a decrease from Day 1 to Day 5 (median estimate = -0.02, 95% HPD: -0.06, -0.004), with no other credible differences across days. Anxious demeanor remained stable, with no credible changes across any test days. Neutral demeanor decreased over time, with lower probabilities on Day 6A relative to Day 1 (median estimate = -0.20, 95% HPD: -0.34, -0.07; Figure 4b). Overall, these results suggest that under the positive cue, dogs increasingly displayed affiliative behaviour and reduced neutral behaviour over time, while aggressive and anxious behaviours remained largely stable. The results of all pairwise contrasts are presented in Table S4.

**c) What is the difference in approachability test demeanor across days within the threatening cue?**

Affiliative demeanor declined over time, with lower probabilities observed on Day 5 (median estimate = -0.17, 95% HPD: -0.29, -0.05) and Day 6B (median estimate = -0.17, 95% HPD: -0.29, -0.06) relative to Day 1. Aggressive demeanor showed credible reductions from Day 1 to later days, including Day 3 (median estimate = -0.04, 95% HPD: -0.09, -0.001), Day 4 (median estimate = -0.04, 95% HPD: -0.10, -0.01), Day 5 (median estimate = -0.05, 95% HPD: -0.10, -0.02), and Day 6B (median estimate = -0.05, 95% HPD: -0.11, -0.01). Anxious demeanor increased from Day 1 to Day 3 (median estimate = 0.05, 95% HPD: 0.001, 0.10). Neutral demeanor increased over time, with higher probabilities on Day 5 (median estimate = 0.23, 95% HPD: 0.09, 0.36) and Day 6B (median estimate = 0.22, 95% HPD: 0.08, 0.34) relative to Day 1. An increase was observed from Day 6A to Day 6B (median estimate = 0.15, 95% HPD: 0.01, 0.29; Figure 4c). Overall, these results suggest that under the threatening cue, dogs showed a shift over time from affiliative and, to a lesser extent, aggressive and anxious behaviours toward increased neutral behaviour with repeated exposure. The results of all pairwise contrasts are presented in Table S5.

**d) What is the difference in approachability test demeanor between male and female dogs?**

Females showed a lower predicted probability for displaying the affiliative demeanor (median difference = -0.12; 95% HPD: -0.20, -0.02) and a higher predicted probability of displaying the neutral demeanor (median difference = 0.13; 95% HPD: 0.04, 0.23) than males (Figure 4d). The results of all pairwise contrasts are presented in Table S6.

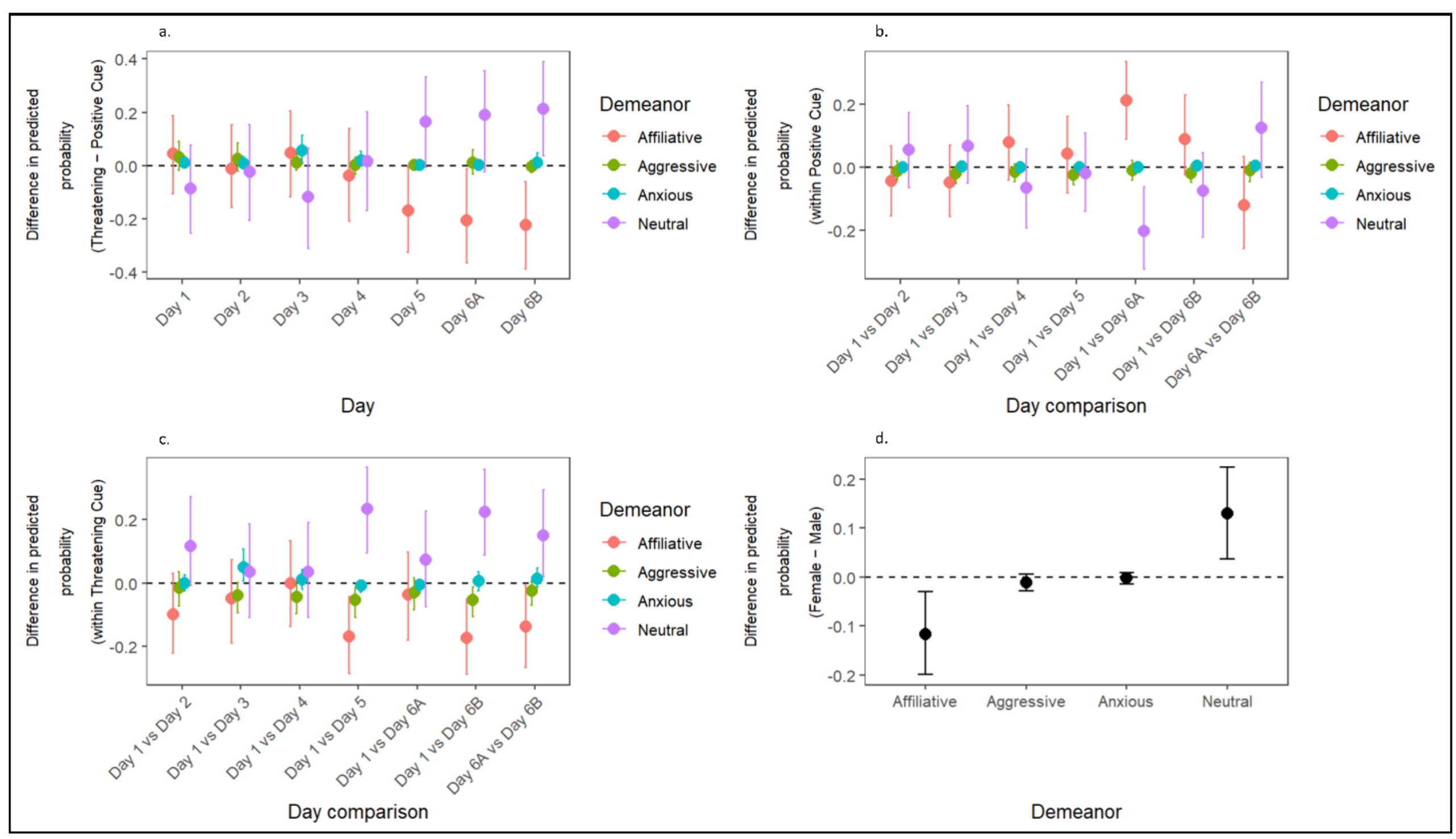

**Figure 4.** Difference in predicted probability for the different demeanors during the approachability test: a) Between-cue contrasts (Threatening - Positive Cue) across days, b) Within-positive cue contrasts between selected day comparisons (relative to Day 1 and between Days 6A and 6B), c) Within-threatening cue contrasts showing analogous day-to-day changes, d) Between-sex contrasts (Female - Male). Points represent posterior medians and error bars indicate 95% HPD. The dashed horizontal line at zero indicates no difference.

**4) What factors affect demeanor during the test phase?**

The model revealed significant effects of day and cue on dogs' demeanor during the approachability test (Table 4).

**Table 4.** Posterior parameter estimates (mean, SE, and 95% credible intervals) from the Bayesian multinomial regression model assessing the effects of day, cue type, sex, and their interaction on demeanor categories (aggressive, anxious and neutral, affiliative) during the test phase, including model diagnostics ($\hat{R}$, Bulk ESS, Tail ESS). "AG", "AN" and "N" represent aggressive, anxious and neutral demeanors compared against the reference demeanor "AF" or affiliative. Parameters marked with an asterisk (*) indicate statistically credible effects where the 95% CI completely excludes zero.

| Parameter | Estimate | Est.Error | l-95% CI | u-95% CI | Rhat | Bulk_ESS | Tail_ESS |
|---|---|---|---|---|---|---|---|
| **muAG_Intercept *** | **-2.81** | **0.55** | **-3.94** | **-1.78** | **1.00** | **11199** | **12861** |
| **muAN_Intercept *** | **-3.71** | **0.65** | **-5.03** | **-2.50** | **1.00** | **16047** | **12124** |
| **muN_Intercept *** | **1.14** | **0.26** | **0.62** | **1.66** | **1.00** | **12648** | **12221** |
| **muAG_CUETHREATENING *** | **1.50** | **0.54** | **0.44** | **2.55** | **1.00** | **14737** | **12710** |
| muAG_DAYDay2 | 0.67 | 0.53 | -0.37 | 1.68 | 1.00 | 18976 | 13998 |
| muAG_DAYDay3 | -0.87 | 0.61 | -2.13 | 0.29 | 1.00 | 22370 | 11922 |
| muAG_DAYDay4 | -0.51 | 0.58 | -1.65 | 0.60 | 1.00 | 23626 | 12919 |
| muAG_DAYDay5 | -0.80 | 0.61 | -2.04 | 0.38 | 1.00 | 20735 | 12293 |
| muAG_SEXM | -0.08 | 0.38 | -0.84 | 0.67 | 1.00 | 21721 | 12963 |
| muAG_CUETHREATENING:DAYDay2 | 0.43 | 0.60 | -0.75 | 1.61 | 1.00 | 18012 | 12898 |
| muAG_CUETHREATENING:DAYDay3 | 0.30 | 0.68 | -1.01 | 1.65 | 1.00 | 21883 | 13021 |
| muAG_CUETHREATENING:DAYDay4 | -0.65 | 0.68 | -1.99 | 0.71 | 1.00 | 23029 | 13034 |
| muAG_CUETHREATENING:DAYDay5 | 0.32 | 0.69 | -1.03 | 1.65 | 1.00 | 21952 | 13981 |
| **muAN_CUETHREATENING *** | **1.50** | **0.62** | **0.27** | **2.70** | **1.00** | **16137** | **12923** |
| muAN_DAYDay2 | -0.06 | 0.66 | -1.36 | 1.19 | 1.00 | 21239 | 12386 |
| muAN_DAYDay3 | 0.08 | 0.63 | -1.16 | 1.28 | 1.00 | 22096 | 13435 |
| muAN_DAYDay4 | -0.45 | 0.66 | -1.77 | 0.80 | 1.00 | 20068 | 12194 |
| muAN_DAYDay5 | -0.44 | 0.67 | -1.78 | 0.83 | 1.00 | 21784 | 12042 |
| muAN_SEXM | -0.01 | 0.44 | -0.87 | 0.87 | 1.00 | 24814 | 12435 |
| muAN_CUETHREATENING:DAYDay2 | 0.68 | 0.70 | -0.69 | 2.06 | 1.00 | 21144 | 12832 |
| muAN_CUETHREATENING:DAYDay3 | 0.38 | 0.67 | -0.92 | 1.70 | 1.00 | 19962 | 13226 |
| muAN_CUETHREATENING:DAYDay4 | 0.29 | 0.70 | -1.09 | 1.66 | 1.00 | 19519 | 12671 |
| muAN_CUETHREATENING:DAYDay5 | 0.20 | 0.73 | -1.23 | 1.63 | 1.00 | 21618 | 12805 |
| muN_CUETHREATENING | 0.04 | 0.33 | -0.61 | 0.69 | 1.00 | 11189 | 12276 |
| muN_DAYDay2 | 0.42 | 0.31 | -0.18 | 1.04 | 1.00 | 15379 | 13198 |
| muN_DAYDay3 | 0.09 | 0.29 | -0.47 | 0.67 | 1.00 | 14344 | 11933 |
| muN_DAYDay4 | -0.03 | 0.29 | -0.60 | 0.53 | 1.00 | 13854 | 13131 |
| muN_DAYDay5 | 0.26 | 0.29 | -0.32 | 0.84 | 1.00 | 14702 | 12676 |
| muN_SEXM | -0.15 | 0.20 | -0.55 | 0.24 | 1.00 | 17634 | 13050 |
| muN_CUETHREATENING:DAYDay2 | 0.35 | 0.44 | -0.52 | 1.22 | 1.00 | 15644 | 12763 |
| muN_CUETHREATENING:DAYDay3 | 0.13 | 0.41 | -0.68 | 0.95 | 1.00 | 14614 | 12173 |
| muN_CUETHREATENING:DAYDay4 | 0.34 | 0.41 | -0.46 | 1.14 | 1.00 | 14477 | 12756 |
| muN_CUETHREATENING:DAYDay5 | 0.52 | 0.43 | -0.33 | 1.37 | 1.00 | 14741 | 12606 |

The results of pairwise contrasts are described below:

**a) What is the difference in test phase demeanor within day between the positive and threatening cues?**

Aggressive demeanor was consistently higher under the threatening cue compared to the positive cue on Day 1 (median estimate = 0.04, 95% HPD: 0.003, 0.10) and Day 2 (median estimate = 0.07, 95% HPD: 0.01, 0.15; Figure 5a). The results of all pairwise contrasts are presented in Table S7.

**b) What is the difference in test phase demeanor across days within the positive cue?**

Estimated marginal contrasts of predicted probabilities within the positive cue condition indicated no credible changes in any of the demeanour across test days (Figure 5b). The results of all pairwise contrasts are presented in Table S8.

**c) What is the difference in test phase demeanor across days within the threatening cue?**

Affiliative demeanor was lower on Day 2 (median estimate = -0.11, 95% HPD: -0.21, -0.02) and Day 5 (median estimate = -0.10, 95% HPD: -0.20, -0.005) relative to Day 1. Aggressive demeanor showed credible decreases over time, including lower probabilities on Day 4 relative to Day 1 (median estimate = -0.04, 95% HPD: -0.10, -0.003). Anxious demeanor showed no credible changes across any test days. Neutral demeanor increased from Day 1 to Day 5 (median estimate = 0.15, 95% HPD: 0.04, 0.27; Figure 5c). Overall, these results suggest that under the threatening cue, early changes were characterised by a reduction in affiliative and aggressive behaviours alongside an increase in neutral behaviour, while anxious behaviour remained stable. The results of all pairwise contrasts are presented in Table S9.

**d) What is the difference in test phase demeanor between male and female dogs?**

Estimated marginal contrasts of predicted probabilities indicated no credible differences between sexes across demeanour categories and test days (Figure 5d). The results of all pairwise contrasts are presented in Table S10.

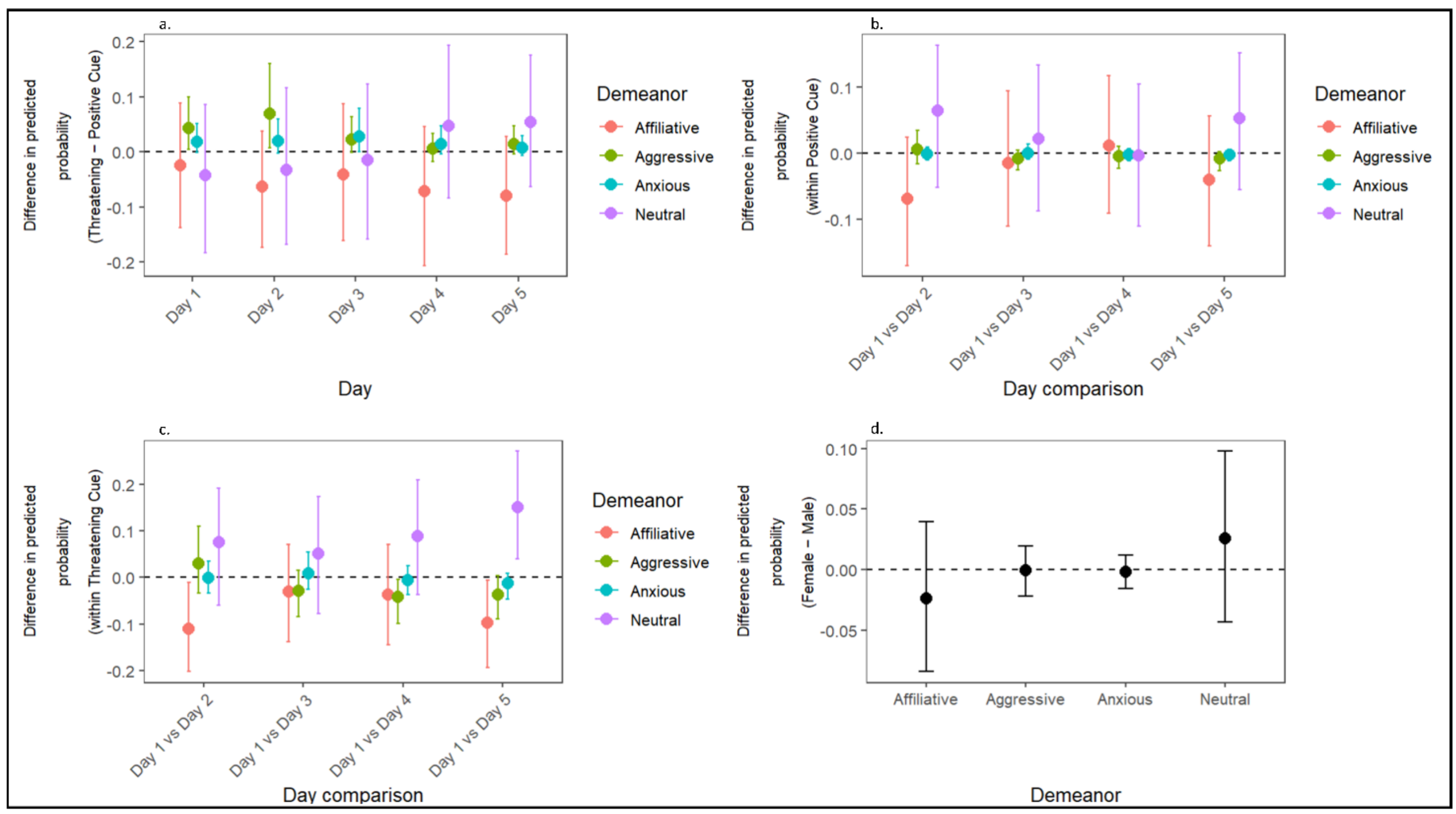

**Figure 5.** Difference in predicted probability for the different demeanors during the test phase: a) Between-cue contrasts (Threatening - Positive cue) across days, b) Within-positive cue contrasts between selected day comparisons (relative to Day 1), c) Within-threatening cue contrasts showing analogous day-to-day changes,

d) Between-sex contrasts (Female - Male). Points represent posterior medians and error bars indicate 95% HPD. The dashed horizontal line at zero indicates no difference.

**Discussion**

The present study investigated how positive and threatening human cues differentially influence sociability in Indian free-ranging dogs, and whether these effects generalize across unfamiliar individuals. Free-ranging dogs in India live in close association with humans (Bhattacharjee and Bhadra, 2020) and are routinely exposed to both affiliative and aversive interactions, some of which can be severe or even lethal (Paul et al., 2016; Singh, 2025). Such experiences are likely to play a critical role in shaping their behavioural strategies toward humans (Bhattacharjee et al., 2017). The ability to evaluate and learn from human interactions is therefore highly adaptive, enabling dogs to maximize benefits from positive interactions while minimizing risks associated with potential threats (Rault et al., 2020).

While one-time tests have been carried out earlier to understand the response of FRDs to positive and threatening cues by unfamiliar humans, this study design enabled us to understand how they adjust their responses to repeated exposure to a cue. On Day 1, groups exposed to the threatening cue exhibited a higher proportion of approach behaviour compared to those exposed to the positive cue. This pattern appears to be driven by the distribution of extreme responses, with some threatening-cue groups showing complete approach and some positive-cue groups showing complete non-approach. Since the cue was provided after the first approach on Day 1, this response was independent of the cue, and as cue selection was randomized prior to the experiment, this differential approach did not influence the cue provided to the group.

Unexpectedly, groups exposed to the positive cue showed increased approach behaviour on Day 6A when tested with an unfamiliar experimenter, but not on other days with the familiar individual. This counterintuitive pattern may reflect differences in dogs' responses to individual humans rather than a simple effect of cue valence. Previous work has shown that dogs vary in their approachability toward different experimenters (Nandi et al., 2026), which may explain this result. Alternatively, repeated exposure to positive cues (e.g., food rewards) may promote a generalized increase in trust toward humans (Rault et al., 2020), leading to enhanced approach behaviour toward unfamiliar individuals, even if this effect is not consistently maintained across contexts.

In contrast, dogs exposed to the threatening cue showed reduced approach behaviour on later days (Day 5 and Day 6B), indicating that repeated exposure to aversive cues decreases approachability over time. Notably, a clear negativity bias was not observed: dogs exposed to the threatening cue did not show reduced approach toward the unfamiliar experimenter on Day 6A. This suggests that dogs may maintain an initial baseline level of trust toward humans, updating their behaviour gradually based on accumulated experience rather than immediately generalizing negative associations.

Patterns in approach latency further supported these interpretations. On Day 1, dogs exposed to the positive cue showed higher approach latency than those exposed to the threatening cue, likely reflecting the inherent extreme response patterns observed in approachability. However, from Day 4 onward (Days 4, 5, and 6B), dogs exposed to the threatening cue exhibited longer latencies compared to those exposed to the positive cue. No difference in approach latency was observed for dogs exposed to the threatening cue between Day 1 and Day 6A, further supporting the absence of generalization for this cue. Additionally, no difference in approach latency was found between the positive and threatening cue groups on Day 6A, indicating that prior cue valence did not influence motivation to approach an unfamiliar person. Overall, these findings indicate that cue valence affects not only the likelihood of approach but also the underlying motivation, with latency serving as an index of hesitation or uncertainty.

Within the positive cue condition, approach latencies decreased significantly on Days 4, 5, and 6B, suggesting increased sociability or motivation with repeated exposure. No such change was observed on Day 6A, in spite of the fact that dogs showed increased approach on Day 6A, reinforcing the idea that although positive reinforcement increases sociability towards the unfamiliar person, dogs still approach them with hesitation. In contrast, dogs exposed to the threatening cue showed increased latencies from Day 2 onward relative to Day 1, indicating a progressive and faster decline in approach motivation in the groups exposed to the threatening cue compared to those exposed to the positive cue. This pattern appears to be driven both by increased hesitation among dogs that approached and by a growing proportion of dogs that did not approach at all, as reflected in the increasing number of censored observations on later days.

Demeanor during the approachability test also differed across conditions. Dogs exposed to the positive cue exhibited significantly lower anxious demeanor than those exposed to the threatening cue on Days 3 and 4, suggesting that repeated exposure to threatening cues initially elevates anxiety but that this effect diminishes over time. Also, no difference in anxious demeanor was observed on Day 6A, again reflecting that anxious demeanors do not generalize to unfamiliar individuals. Affiliative demeanor was higher in the positive cue groups on Days 5, 6A, and 6B, indicating that once affiliative responses develop, they persist and generalize even to unfamiliar individuals. Neutral behaviour was higher in the threatening cue groups on Day 6B, suggesting a slower emergence of behavioural differences that do not generalize to unfamiliar contexts. Sex differences were also evident. Female dogs displayed higher neutral and lower affiliative demeanor across days, indicating greater caution compared to males, while anxious and aggressive behaviours did not differ between sexes.

During the test phase, aggressive demeanor was higher in threatening cue groups on the initial days (Days 1 and 2), but this difference disappeared thereafter, possibly reflecting habituation to the experimental stimulus (e.g., the stick) when no actual harm was experienced. Other demeanors did not differ between cue conditions. No sex differences or within-positive cue differences were observed during this phase. However, within the threatening cue condition, affiliative demeanor decreased on Days 2 and 5 relative to Day 1, aggressive demeanor decreased by Day 4, and neutral demeanor increased by Day 5, indicating a gradual shift in behavioural responses from aggressive to neutral over repeated exposures.

Overall, our results indicate that free-ranging dogs partially generalize positive cues across individuals, which was evident from the change in approach proportion but was not reflected in the change in approach latency, whereas threatening cues are not generalized in the same way. Specifically, while exposure to threatening cues does not reduce dogs' willingness to approach unfamiliar humans, it does increase their approach latency, indicating greater caution. In practical terms, receiving food or affiliative signals from one human appears to enhance dogs' overall trust and sociability toward unfamiliar individuals. In contrast, exposure to a threatening interaction from one person does not broadly suppress approach behaviour toward other humans. This asymmetry in learning likely reflects an adaptive strategy in environments where human-dog interactions are highly variable and unpredictable. By maintaining a baseline willingness to approach while simultaneously increasing caution in response to potential threats, free-ranging dogs may balance the need to exploit resource opportunities with the need to minimize risk.

Our findings also highlight the dynamic interplay between behavioural plasticity and stable individual tendencies in free-ranging dogs. Several individuals who initially failed to approach on Day 1 began to approach by Day 6A following repeated exposure to positive cues (e.g., food rewards), whereas some individuals who initially approached ceased to do so after repeated exposure to threatening cues. These patterns provide clear evidence of experience-dependent behavioural modification, consistent with operant learning processes.

The results of this study have important implications for human-dog coexistence in urban environments. By elucidating how free-ranging dogs respond to and learn from different types of human interactions, these findings can inform public awareness initiatives, improve the design and welfare outcomes of stray dog management programs, and contribute to reducing human-animal conflict in the Anthropocene.

**Acknowledgements**

The authors gratefully acknowledge the Indian Institute of Science Education and Research Kolkata for providing infrastructural support. They also thank Arpan Bhattacharyya for creating the experimental illustration and Imran Mondal for assistance with the fieldwork.

**Funding**

SN would like to thank the Department of Science and Technology, India for providing her doctoral fellowship. The project was partially funded by the Janaki Ammal Award grant BT/HRD/NBA-NWB/39/2020-21 (YC-1), to AB by the Department of Biotechnology, India.

**CRediT authorship contribution statement**

Srijaya Nandi: Conceptualization, Methodology, Investigation, Data curation, Software, Formal analysis, Writing - Original Draft, Visualization, Project administration. Dipanjan Roy: Investigation, Data curation, Validation. Abantika Adak: Investigation, Data curation, Validation. Rohan Sarkar: Software, Writing - Review & Editing. Anindita Bhadra: Conceptualization, Methodology, Resources, Writing - Review & Editing, Supervision, Project administration, Funding acquisition.

**Data availability statement**

Data supporting the results will be archived.

**Conflict of Interest Information**

The authors declare no conflict of interest.

**Supplementary information**

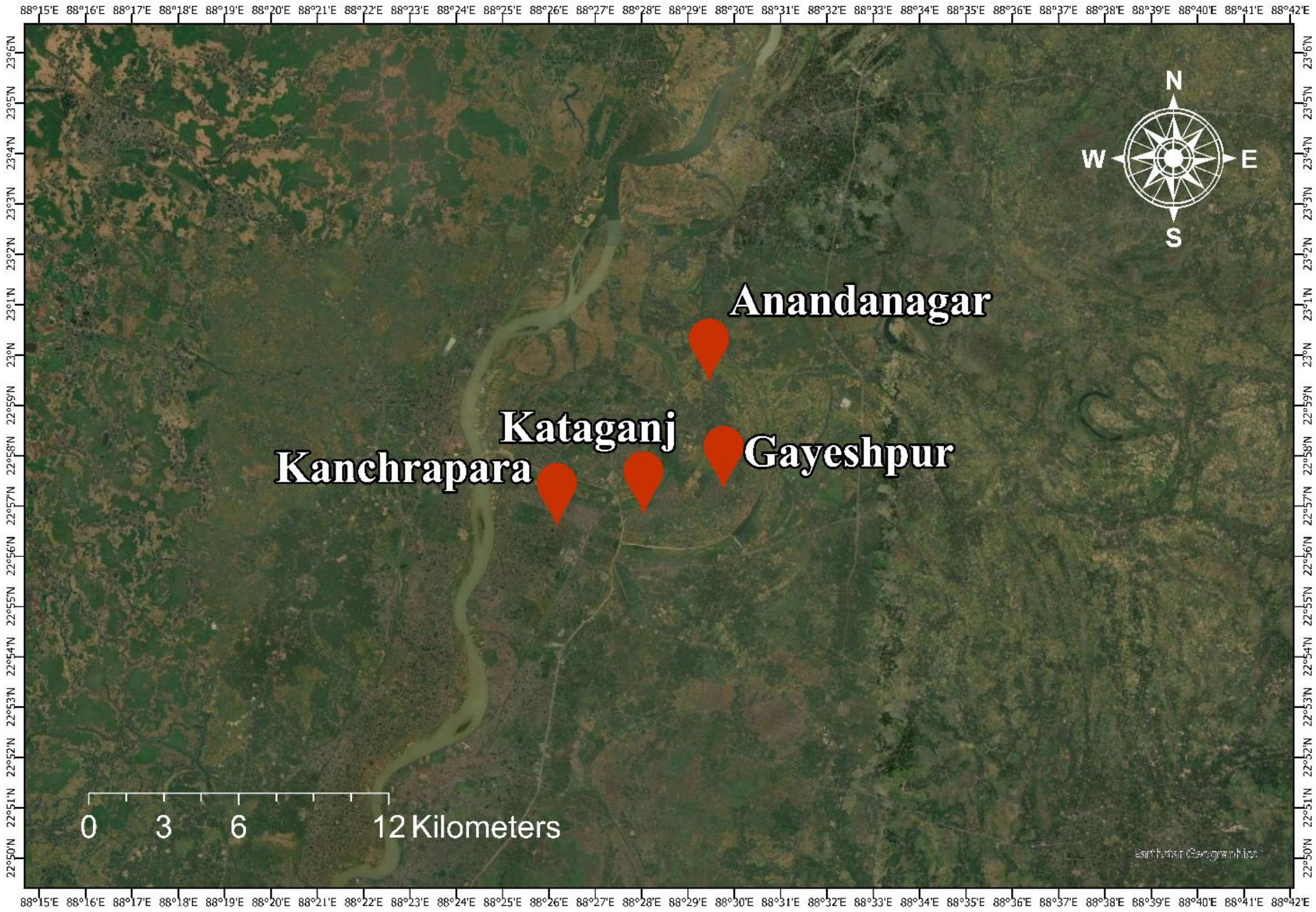

**Figure S1.** Study sites

**Table S1.** Posterior summary of pairwise contrast comparisons for approach proportion during the approachability test across days and cue conditions. Estimates represent the median posterior differences between conditions. Values in parentheses denote the 95% Highest Posterior Density (HPD) intervals. Pairwise comparisons marked with an asterisk (*) indicate statistically credible differences where the 95% HPD interval completely excludes zero.

| Contrast Comparison | Estimate | Lower HPD (95%) | Upper HPD (95%) |
|---|---|---|---|
| Day 1 POSITIVE - Day 2 POSITIVE | -0.052484 | -0.16385 | 0.06396 |
| Day 1 POSITIVE - Day 3 POSITIVE | -0.035061 | -0.12719 | 0.05487 |
| Day 1 POSITIVE - Day 4 POSITIVE | -0.037192 | -0.13829 | 0.06239 |
| Day 1 POSITIVE - Day 5 POSITIVE | -0.056893 | -0.16098 | 0.04836 |
| **Day 1 POSITIVE - Day 6A POSITIVE *** | **-0.119523** | **-0.22675** | **-0.01584** |
| Day 1 POSITIVE - Day 6B POSITIVE | -0.078090 | -0.17651 | 0.02669 |
| **Day 1 POSITIVE - Day 1 THREATENING *** | **-0.103875** | **-0.20118** | **-0.00501** |
| Day 1 POSITIVE - Day 2 THREATENING | -0.037645 | -0.16121 | 0.07916 |
| Day 1 POSITIVE - Day 3 THREATENING | -0.010869 | -0.14183 | 0.10501 |
| Day 1 POSITIVE - Day 4 THREATENING | -0.060329 | -0.16412 | 0.04186 |
| Day 1 POSITIVE - Day 5 THREATENING | 0.001302 | -0.10863 | 0.11009 |
| Day 1 POSITIVE - Day 6A THREATENING | -0.061864 | -0.18451 | 0.05938 |
| Day 1 POSITIVE - Day 6B THREATENING | 0.031250 | -0.08295 | 0.14900 |
| Day 2 POSITIVE - Day 3 POSITIVE | 0.017803 | -0.09978 | 0.12419 |
| Day 2 POSITIVE - Day 4 POSITIVE | 0.015320 | -0.10301 | 0.13870 |
| Day 2 POSITIVE - Day 5 POSITIVE | -0.004379 | -0.12329 | 0.12048 |
| Day 2 POSITIVE - Day 6A POSITIVE | -0.066215 | -0.19167 | 0.05584 |
| Day 2 POSITIVE - Day 6B POSITIVE | -0.024447 | -0.14300 | 0.09961 |
| Day 2 POSITIVE - Day 1 THREATENING | -0.050895 | -0.17589 | 0.06948 |
| Day 2 POSITIVE - Day 2 THREATENING | 0.014799 | -0.12231 | 0.14524 |
| Day 2 POSITIVE - Day 3 THREATENING | 0.041045 | -0.09704 | 0.18241 |
| Day 2 POSITIVE - Day 4 THREATENING | -0.007254 | -1.12958 | 0.11579 |
| Day 2 POSITIVE - Day 5 THREATENING | 0.053655 | -0.07416 | 0.18326 |
| Day 2 POSITIVE - Day 6A THREATENING | -0.008629 | -0.14777 | 0.12400 |
| Day 2 POSITIVE - Day 6B THREATENING | 0.084253 | -0.04652 | 0.21716 |
| Day 3 POSITIVE - Day 4 POSITIVE | -0.001924 | -0.09751 | 0.09498 |
| Day 3 POSITIVE - Day 5 POSITIVE | -0.022481 | -0.12388 | 0.07495 |
| Day 3 POSITIVE - Day 6A POSITIVE | -0.084026 | -0.18301 | 0.01754 |
| Day 3 POSITIVE - Day 6B POSITIVE | -0.042670 | -0.13723 | 0.05636 |
| Day 3 POSITIVE - Day 1 THREATENING | -0.068998 | -0.16616 | 0.02851 |
| Day 3 POSITIVE - Day 2 THREATENING | -0.002753 | -0.11543 | 0.11117 |
| Day 3 POSITIVE - Day 3 THREATENING | 0.023665 | -0.09066 | 0.14187 |
| Day 3 POSITIVE - Day 4 THREATENING | -0.025733 | -0.12119 | 0.07277 |
| Day 3 POSITIVE - Day 5 THREATENING | 0.037142 | -0.06735 | 0.14256 |
| Day 3 POSITIVE - Day 6A THREATENING | -0.026555 | -0.14068 | 0.08876 |
| Day 3 POSITIVE - Day 6B THREATENING | 0.066320 | -0.04547 | 0.17482 |
| Day 4 POSITIVE - Day 5 POSITIVE | -0.019903 | -0.12614 | 0.08872 |
| Day 4 POSITIVE - Day 6A POSITIVE | -0.081698 | -0.19507 | 0.02507 |
| Day 4 POSITIVE - Day 6B POSITIVE | -0.040619 | -0.14507 | 0.06759 |
| Day 4 POSITIVE - Day 1 THREATENING | -0.066382 | -0.17369 | 0.03795 |
| Day 4 POSITIVE - Day 2 THREATENING | -0.000642 | -0.12137 | 0.12177 |
| Day 4 POSITIVE - Day 3 THREATENING | 0.026150 | -0.09801 | 0.15338 |
| Day 4 THREATENING - Day 4 THREATENING | -0.023336 | -0.12880 | 0.07990 |
| Day 4 POSITIVE - Day 5 THREATENING | 0.038826 | -0.07881 | 0.14775 |
| Day 4 POSITIVE - Day 6A THREATENING | -0.024682 | -0.14103 | 0.10557 |
| Day 4 POSITIVE - Day 6B THREATENING | 0.068187 | -0.05002 | 0.18705 |
| Day 5 POSITIVE - Day 6A POSITIVE | -0.061337 | -0.17596 | 0.04983 |
| Day 5 POSITIVE - Day 6B POSITIVE | -0.020465 | -0.12748 | 0.09152 |
| Day 5 POSITIVE - Day 1 THREATENING | -0.046089 | -0.15440 | 0.06833 |
| Day 5 POSITIVE - Day 2 THREATENING | 0.019540 | -0.10815 | 0.14483 |
| Day 5 POSITIVE - Day 3 THREATENING | 0.045623 | -0.08187 | 0.17389 |
| Day 5 POSITIVE - Day 4 THREATENING | -0.002709 | -0.11158 | 0.11180 |
| Day 5 POSITIVE - Day 5 THREATENING | 0.058694 | -0.05571 | 0.17714 |
| Day 5 POSITIVE - Day 6A THREATENING | -0.004594 | -0.12950 | 0.12559 |
| Day 5 POSITIVE - Day 6B THREATENING | 0.088551 | -0.03307 | 0.20994 |
| Day 6A POSITIVE - Day 6B POSITIVE | 0.041711 | -0.06879 | 0.15368 |
| Day 6A POSITIVE - Day 1 THREATENING | 0.014914 | -0.09507 | 0.12718 |
| Day 6A POSITIVE - Day 2 THREATENING | 0.081295 | -0.05356 | 0.20421 |
| Day 6A POSITIVE - Day 3 THREATENING | 0.107882 | -0.02329 | 0.23649 |
| Day 6A POSITIVE - Day 4 THREATENING | 0.058660 | -0.05018 | 0.17117 |
| **Day 6A POSITIVE - Day 5 THREATENING *** | **0.120840** | **0.00285** | **0.23847** |
| Day 6A POSITIVE - Day 6A THREATENING | 0.057060 | -0.06654 | 0.18195 |
| **Day 6A POSITIVE - Day 6B THREATENING *** | **0.150670** | **0.02650** | **0.27041** |
| Day 6B POSITIVE - Day 1 THREATENING | -0.026606 | -0.13289 | 0.08228 |
| Day 6B POSITIVE - Day 2 THREATENING | 0.040153 | -0.08349 | 0.15948 |
| Day 6B POSITIVE - Day 3 THREATENING | 0.066790 | -0.06495 | 0.18810 |
| Day 6B POSITIVE - Day 4 THREATENING | 0.017201 | -0.09142 | 0.12361 |
| Day 6B POSITIVE - Day 5 THREATENING | 0.079196 | -0.03787 | 0.19175 |
| Day 6B POSITIVE - Day 6A THREATENING | 0.015835 | -0.10917 | 0.13907 |
| Day 6B POSITIVE - Day 6B THREATENING | 0.108430 | -0.01341 | 0.22010 |
| Day 1 THREATENING - Day 2 THREATENING | 0.066294 | -0.04116 | 0.16993 |
| Day 1 THREATENING - Day 3 THREATENING | 0.093276 | -0.01604 | 0.20142 |
| Day 1 THREATENING - Day 4 THREATENING | 0.043326 | -0.04539 | 0.13169 |
| **Day 1 THREATENING - Day 5 THREATENING *** | **0.105884** | **0.00914** | **0.20120** |
| Day 1 THREATENING - Day 6A THREATENING | 0.042333 | -0.06407 | 0.15091 |
| **Day 1 THREATENING - Day 6B THREATENING *** | **0.135371** | **0.03536** | **0.23764** |
| Day 2 THREATENING - Day 3 THREATENING | 0.026308 | -0.09410 | 0.15421 |
| Day 2 THREATENING - Day 4 THREATENING | -0.022390 | -0.13029 | 0.08301 |
| Day 2 THREATENING - Day 5 THREATENING | 0.039371 | -0.07345 | 0.15361 |
| Day 2 THREATENING - Day 6A THREATENING | -0.024125 | -0.14674 | 0.09873 |
| Day 2 THREATENING - Day 6B THREATENING | 0.069098 | -0.05203 | 0.18600 |
| Day 3 THREATENING - Day 4 THREATENING | -0.049330 | -0.15966 | 0.06167 |
| Day 3 THREATENING - Day 5 THREATENING | 0.012222 | -1.10002 | 0.13194 |
| Day 3 THREATENING - Day 6A THREATENING | -0.050543 | -0.17465 | 0.07313 |
| Day 3 THREATENING - Day 6B THREATENING | 0.042934 | -0.07859 | 0.16324 |
| Day 4 THREATENING - Day 5 THREATENING | 0.061986 | -0.03209 | 0.16013 |
| Day 4 THREATENING - Day 6A THREATENING | -0.001511 | -0.10768 | 0.10971 |
| Day 4 THREATENING - Day 6B THREATENING | 0.092041 | -0.01142 | 0.19207 |
| Day 5 THREATENING - Day 6A THREATENING | -0.062422 | -0.17642 | 0.05515 |
| Day 5 THREATENING - Day 6B THREATENING | 0.029668 | -0.07992 | 0.13816 |
| Day 6A THREATENING - Day 6B THREATENING | 0.093044 | -0.02257 | 0.21296 |

**Table S2.** Posterior summary of pairwise contrast comparisons for approach latency during the approachability test across days and cue conditions. Estimates represent the median posterior

differences between conditions. Values in parentheses denote the 95% Highest Posterior Density (HPD) intervals. Pairwise comparisons marked with an asterisk (*) indicate statistically credible differences where the 95% HPD interval completely excludes zero.

| Contrast Comparison | Estimate | Lower HPD (95%) | Upper HPD (95%) |
|---|---|---|---|
| Day 1 POSITIVE - Day 2 POSITIVE | 0.101522 | -0.40544 | 0.62278 |
| Day 1 POSITIVE - Day 3 POSITIVE | 0.270703 | -0.22190 | 0.77405 |
| **Day 1 POSITIVE - Day 4 POSITIVE *** | **0.824274** | **0.34914** | **1.30488** |
| **Day 1 POSITIVE - Day 5 POSITIVE *** | **0.690945** | **0.21730** | **1.19542** |
| Day 1 POSITIVE - Day 6A POSITIVE | 0.282670 | -0.24085 | 0.76918 |
| **Day 1 POSITIVE - Day 6B POSITIVE *** | **0.504362** | **0.00523** | **0.99809** |
| **Day 1 POSITIVE - Day 1 THREATENING *** | **0.782678** | **0.14207** | **1.43215** |
| Day 1 POSITIVE - Day 2 THREATENING | -0.159279 | -0.89754 | 0.58968 |
| Day 1 POSITIVE - Day 3 THREATENING | -0.390961 | -1.17740 | 0.35018 |
| Day 1 POSITIVE - Day 4 THREATENING | 0.037473 | -0.70272 | 0.76861 |
| Day 1 POSITIVE - Day 5 THREATENING | -0.396813 | -1.15394 | 0.35197 |
| Day 1 POSITIVE - Day 6A THREATENING | 0.142728 | -0.60756 | 0.86312 |
| Day 1 POSITIVE - Day 6B THREATENING | -0.465475 | -1.23757 | 0.28591 |
| Day 2 POSITIVE - Day 3 POSITIVE | 0.165975 | -0.39542 | 0.71293 |
| **Day 2 POSITIVE - Day 4 POSITIVE *** | **0.720761** | **0.17702** | **1.25212** |
| **Day 2 POSITIVE - Day 5 POSITIVE *** | **0.591363** | **0.04516** | **1.13036** |
| Day 2 POSITIVE - Day 6A POSITIVE | 0.176139 | -0.36832 | 0.75927 |
| Day 2 POSITIVE - Day 6B POSITIVE | 0.403844 | -0.14667 | 0.97226 |

| Day 2 POSITIVE - Day 1 THREATENING | 0.685521 | -0.03821 | 1.38742 |
|---|---|---|---|
| Day 2 POSITIVE - Day 2 THREATENING | -0.255791 | -1.01536 | 0.49495 |
| Day 2 POSITIVE - Day 3 THREATENING | -0.491980 | -1.28293 | 0.29493 |
| Day 2 POSITIVE - Day 4 THREATENING | -0.066539 | -0.83504 | 0.69658 |
| Day 2 POSITIVE - Day 5 THREATENING | -0.496537 | -1.29386 | 0.29667 |
| Day 2 POSITIVE - Day 6A THREATENING | 0.041229 | -0.72717 | 0.79335 |
| Day 2 POSITIVE - Day 6B THREATENING | -0.562870 | -1.34809 | 0.24088 |
| **Day 3 POSITIVE - Day 4 POSITIVE *** | **0.554886** | **0.02676** | **1.08228** |
| Day 3 POSITIVE - Day 5 POSITIVE | 0.421274 | -0.10176 | 0.97347 |
| Day 3 POSITIVE - Day 6A POSITIVE | 0.010623 | -0.54792 | 0.55199 |
| Day 3 POSITIVE - Day 6B POSITIVE | 0.236272 | -0.31445 | 0.77166 |
| Day 3 POSITIVE - Day 1 THREATENING | 0.515960 | -0.16015 | 1.25182 |
| Day 3 POSITIVE - Day 2 THREATENING | -0.429511 | -1.17969 | 0.35069 |
| Day 3 POSITIVE - Day 3 THREATENING | -0.661038 | -1.42572 | 0.10067 |
| Day 3 POSITIVE - Day 4 THREATENING | -0.240140 | -0.98075 | 0.53611 |
| Day 3 POSITIVE - Day 5 THREATENING | -0.667501 | -1.44421 | 0.10592 |
| Day 3 POSITIVE - Day 6A THREATENING | -0.130543 | -0.88750 | 0.60938 |
| Day 3 POSITIVE - Day 6B THREATENING | -0.736580 | -1.50622 | 0.06483 |
| Day 4 POSITIVE - Day 5 POSITIVE | -0.133708 | -0.66431 | 0.36922 |
| **Day 4 POSITIVE - Day 6A POSITIVE *** | **-0.543579** | **-1.06049** | **-0.00104** |
| Day 4 POSITIVE - Day 6B POSITIVE | -0.318233 | -0.81956 | 0.20897 |
| Day 4 POSITIVE - Day 1 THREATENING | -0.037640 | -0.74592 | 0.63666 |
| **Day 4 POSITIVE - Day 2 THREATENING *** | **-0.984878** | **-1.68980** | **-0.18977** |
| **Day 4 POSITIVE - Day 3 THREATENING *** | **-1.217757** | **-1.97814** | **-0.46418** |
| **Day 4 POSITIVE - Day 4 THREATENING *** | **-0.791801** | **-1.51579** | **-0.05360** |
| **Day 4 POSITIVE - Day 5 THREATENING *** | **-1.221792** | **-1.96341** | **-0.41890** |
| Day 4 POSITIVE - Day 6A THREATENING | -0.682438 | -1.41444 | 0.05805 |
| **Day 4 POSITIVE - Day 6B THREATENING *** | **-1.293182** | **-2.07342** | **-0.54449** |
| Day 5 POSITIVE - Day 6A POSITIVE | -0.411635 | -0.95812 | 0.10733 |
| Day 5 POSITIVE - Day 6B POSITIVE | -0.182834 | -0.71750 | 0.34728 |
| Day 5 POSITIVE - Day 1 THREATENING | 0.096605 | -0.61038 | 0.78607 |
| **Day 5 POSITIVE - Day 2 THREATENING *** | **-0.848931** | **-1.63074** | **-0.10848** |
| **Day 5 POSITIVE - Day 3 THREATENING *** | **-1.080485** | **-1.85618** | **-0.30470** |
| Day 5 POSITIVE - Day 4 THREATENING | -0.660410 | -1.41804 | 0.08940 |
| **Day 5 POSITIVE - Day 5 THREATENING *** | **-1.088421** | **-1.86948** | **-0.34015** |
| Day 5 POSITIVE - Day 6A THREATENING | -0.548621 | -1.28906 | 0.20478 |
| **Day 5 POSITIVE - Day 6B THREATENING *** | **-1.154310** | **-1.92587** | **-0.37218** |
| Day 6A POSITIVE - Day 6B POSITIVE | 0.225795 | -0.29775 | 0.77690 |
| Day 6A POSITIVE - Day 1 THREATENING | 0.506624 | -0.18442 | 1.23304 |
| Day 6A POSITIVE - Day 2 THREATENING | -0.440804 | -1.23285 | 0.29710 |
| Day 6A POSITIVE - Day 3 THREATENING | -0.675671 | -1.42796 | 0.14036 |
| Day 6A POSITIVE - Day 4 THREATENING | -0.248398 | -1.03575 | 0.49551 |
| Day 6A POSITIVE - Day 5 THREATENING | -0.675539 | -1.46081 | 0.08798 |
| Day 6A POSITIVE - Day 6A THREATENING | -0.141014 | -0.88281 | 0.60633 |
| Day 6A POSITIVE - Day 6B THREATENING | -0.747212 | -1.53137 | 0.03216 |
| Day 6B POSITIVE - Day 1 THREATENING | 0.277290 | -0.44553 | 0.96896 |
| Day 6B POSITIVE - Day 2 THREATENING | -0.663944 | -1.43126 | 0.09935 |
| **Day 6B POSITIVE - Day 3 THREATENING *** | **-0.899006** | **-1.67068** | **-0.12059** |
| Day 6B POSITIVE - Day 4 THREATENING | -0.472368 | -1.22991 | 0.28117 |
| **Day 6B POSITIVE - Day 5 THREATENING *** | **-0.900902** | **-1.68011** | **-0.12528** |
| Day 6B POSITIVE - Day 6A THREATENING | -0.365143 | -1.12342 | 0.37196 |
| **Day 6B POSITIVE - Day 6B THREATENING *** | **-0.974962** | **-1.74192** | **-0.18725** |
| **Day 1 THREATENING - Day 2 THREATENING *** | **-0.945099** | **-1.48150** | **-0.38751** |
| **Day 1 THREATENING - Day 3 THREATENING *** | **-1.175748** | **-1.73881** | **-0.62191** |
| **Day 1 THREATENING - Day 4 THREATENING *** | **-0.749555** | **-1.27922** | **-0.20673** |
| **Day 1 THREATENING - Day 5 THREATENING *** | **-1.180319** | **-1.75990** | **-0.60727** |
| **Day 1 THREATENING - Day 6A THREATENING *** | **-0.641957** | **-1.15184** | **-0.09675** |
| **Day 1 THREATENING - Day 6B THREATENING *** | **-1.254008** | **-1.81702** | **-0.69137** |
| Day 2 THREATENING - Day 3 THREATENING | -0.234573 | -0.84974 | 0.40940 |
| Day 2 THREATENING - Day 4 THREATENING | 0.195832 | -0.42173 | 0.78507 |
| Day 2 THREATENING - Day 5 THREATENING | -0.234957 | -0.89286 | 0.39111 |
| Day 2 THREATENING - Day 6A THREATENING | 0.296949 | -0.30294 | 0.88869 |
| Day 2 THREATENING - Day 6B THREATENING | -0.304401 | -0.94792 | 0.33303 |
| Day 3 THREATENING - Day 4 THREATENING | 0.427156 | -0.18307 | 1.06275 |
| Day 3 THREATENING - Day 5 THREATENING | -0.000731 | -0.62104 | 0.66062 |
| Day 3 THREATENING - Day 6A THREATENING | 0.535445 | -0.07923 | 1.14240 |
| Day 3 THREATENING - Day 6B THREATENING | -0.073837 | -0.71648 | 0.56702 |
| Day 4 THREATENING - Day 5 THREATENING | -0.426777 | -1.06821 | 0.17543 |
| Day 4 THREATENING - Day 6A THREATENING | 0.108955 | -0.47887 | 0.71271 |
| Day 4 THREATENING - Day 6B THREATENING | -0.496742 | -1.14634 | 0.10076 |
| Day 5 THREATENING - Day 6A THREATENING | 0.534989 | -0.07204 | 1.17412 |
| Day 5 THREATENING - Day 6B THREATENING | -0.073607 | -0.74325 | 0.55924 |
| Day 6A THREATENING - Day 6B THREATENING | -0.606338 | -1.23536 | 0.01043 |

**Table S3.** Posterior summary of pairwise contrast comparisons for demeanor during the approachability test between cue conditions. Estimates represent the median posterior differences between conditions. Values in parentheses denote the 95% Highest Posterior Density (HPD) intervals. Pairwise comparisons marked with an asterisk (*) indicate statistically credible differences where the 95% HPD interval completely excludes zero.

| Day | Demeanor | Median | Lower HPD | Upper HPD |
|---|---|---|---|---|
| Day 1 | AF | 0.041279 | -0.09291 | 0.191026 |
| Day 1 | AG | 0.028854 | -0.02289 | 0.089497 |
| Day 1 | AN | 0.010239 | -0.00079 | 0.029749 |
| Day 1 | N | -0.08287 | -0.24906 | 0.079588 |
| Day 2 | AF | -0.01293 | -0.1871 | 0.136402 |
| Day 2 | AG | 0.024106 | -0.02173 | 0.088593 |
| Day 2 | AN | 0.008256 | -0.00889 | 0.037093 |
| Day 2 | N | -0.02527 | -0.20526 | 0.158815 |
| Day 3 | AF | 0.044902 | -0.12421 | 0.204716 |
| Day 3 | AG | 0.009243 | -0.01807 | 0.04638 |
| **Day 3** | **AN *** | **0.056973** | **0.014421** | **0.119974** |
| Day 3 | N | -0.11743 | -0.29291 | 0.081312 |
| Day 4 | AF | -0.03584 | -0.22927 | 0.12608 |
| Day 4 | AG | 0.00054 | -0.02858 | 0.032001 |
| **Day 4** | **AN *** | **0.019838** | **4.64E-05** | **0.056248** |
| Day 4 | N | 0.014914 | -0.17126 | 0.208673 |
| **Day 5** | **AF *** | **-0.1706** | **-0.31297** | **-0.00435** |
| Day 5 | AG | 0.000928 | -0.01789 | 0.022333 |
| Day 5 | AN | 0.002121 | -0.00757 | 0.019857 |
| Day 5 | N | 0.165069 | -0.01507 | 0.313731 |
| **Day 6A** | **AF *** | **-0.20973** | **-0.38801** | **-0.02736** |
| Day 6A | AG | 0.009003 | -0.03576 | 0.05867 |
| Day 6A | AN | 0.002507 | -0.00925 | 0.022063 |
| Day 6A | N | 0.194284 | -0.00534 | 0.382884 |
| **Day 6B** | **AF *** | **-0.22451** | **-0.40289** | **-0.07698** |
| Day 6B | AG | -0.00477 | -0.03118 | 0.016777 |
| Day 6B | AN | 0.012139 | -0.01075 | 0.043846 |
| **Day 6B** | **N *** | **0.214059** | **0.063187** | **0.412541** |

**Table S4.** Posterior summary of pairwise contrast comparisons for demeanor during the approachability test across days within the threatening cue condition. Estimates represent the median posterior differences between conditions. Values in parentheses denote the 95% Highest Posterior Density (HPD) intervals. Pairwise comparisons marked with an asterisk (*) indicate statistically credible differences where the 95% HPD interval completely excludes zero.

| Demeanor | Day (A) | Day (B) | Median | Lower HPD | Upper HPD |
|---|---|---|---|---|---|
| AF | Day 1 | Day 2 | -0.1011 | -0.22261 | 0.029962 |
| AF | Day 1 | Day 3 | -0.04426 | -0.17625 | 0.093028 |
| AF | Day 1 | Day 4 | -0.00301 | -0.13363 | 0.142132 |
| **AF *** | **Day 1** | **Day 5** | **-0.16705** | **-0.28769** | **-0.04648** |
| AF | Day 1 | Day 6A | -0.03838 | -0.17121 | 0.095321 |
| **AF *** | **Day 1** | **Day 6B** | **-0.17359** | **-0.29332** | **-0.05801** |
| AF | Day 2 | Day 3 | 0.055742 | -0.07646 | 0.208449 |
| AF | Day 2 | Day 4 | 0.099622 | -0.04704 | 0.232381 |
| AF | Day 2 | Day 5 | -0.06291 | -0.2071 | 0.054198 |
| AF | Day 2 | Day 6A | 0.064628 | -0.07946 | 0.201834 |
| AF | Day 2 | Day 6B | -0.07129 | -0.20055 | 0.048749 |
| AF | Day 3 | Day 4 | 0.041872 | -0.10341 | 0.189357 |
| AF | Day 3 | Day 5 | -0.12167 | -0.25518 | 0.007637 |
| AF | Day 3 | Day 6A | 0.006507 | -0.14505 | 0.157863 |
| AF | Day 3 | Day 6B | -0.12763 | -0.26223 | 0.006348 |
| **AF *** | **Day 4** | **Day 5** | **-0.16352** | **-0.29373** | **-0.01681** |
| AF | Day 4 | Day 6A | -0.03458 | -0.19551 | 0.112948 |
| **AF *** | **Day 4** | **Day 6B** | **-0.17112** | **-0.31777** | **-0.04437** |
| AF | Day 5 | Day 6A | 0.128796 | -0.01024 | 0.262424 |
| AF | Day 5 | Day 6B | -0.00581 | -0.12897 | 0.10423 |
| AF | Day 6A | Day 6B | -0.13513 | -0.26186 | 0.00114 |
| AG | Day 1 | Day 2 | -0.01478 | -0.0704 | 0.04091 |
| **AG *** | **Day 1** | **Day 3** | **-0.03914** | **-0.09331** | **-0.00151** |
| **AG *** | **Day 1** | **Day 4** | **-0.04392** | **-0.09735** | **-0.00731** |
| **AG *** | **Day 1** | **Day 5** | **-0.05215** | **-0.10583** | **-0.01586** |
| AG | Day 1 | Day 6A | -0.02729 | -0.08099 | 0.016186 |
| **AG *** | **Day 1** | **Day 6B** | **-0.05372** | **-0.10621** | **-0.01251** |
| AG | Day 2 | Day 3 | -0.0235 | -0.07823 | 0.021014 |
| AG | Day 2 | Day 4 | -0.02726 | -0.08333 | 0.011273 |
| **AG *** | **Day 2** | **Day 5** | **-0.03574** | **-0.09379** | **-0.00299** |
| AG | Day 2 | Day 6A | -0.01252 | -0.07147 | 0.032524 |
| **AG *** | **Day 2** | **Day 6B** | **-0.03637** | **-0.09399** | **-0.00237** |
| AG | Day 3 | Day 4 | -0.00431 | -0.03566 | 0.025831 |
| AG | Day 3 | Day 5 | -0.01154 | -0.04662 | 0.010661 |
| AG | Day 3 | Day 6A | 0.009935 | -0.0301 | 0.052431 |
| AG | Day 3 | Day 6B | -0.01264 | -0.04657 | 0.009093 |
| AG | Day 4 | Day 5 | -0.00761 | -0.03683 | 0.013968 |
| AG | Day 4 | Day 6A | 0.013981 | -0.01686 | 0.058549 |
| AG | Day 4 | Day 6B | -0.00837 | -0.03886 | 0.012512 |
| AG | Day 5 | Day 6A | 0.022464 | -0.00393 | 0.067058 |
| AG | Day 5 | Day 6B | -0.00075 | -0.02077 | 0.019741 |
| AG | Day 6A | Day 6B | -0.02325 | -0.068 | 0.003878 |
| AN | Day 1 | Day 2 | -0.00081 | -0.0284 | 0.026064 |
| **AN *** | **Day 1** | **Day 3** | **0.049802** | **0.00066** | **0.105654** |
| AN | Day 1 | Day 4 | 0.010017 | -0.01679 | 0.046833 |
| AN | Day 1 | Day 5 | -0.00848 | -0.03082 | 0.009953 |
| AN | Day 1 | Day 6A | -0.00701 | -0.03039 | 0.011529 |
| AN | Day 1 | Day 6B | 0.00545 | -0.02068 | 0.038717 |
| **AN *** | **Day 2** | **Day 3** | **0.050326** | **0.003945** | **0.114973** |
| AN | Day 2 | Day 4 | 0.011144 | -0.02741 | 0.048873 |
| AN | Day 2 | Day 5 | -0.00714 | -0.03479 | 0.016783 |
| AN | Day 2 | Day 6A | -0.00596 | -0.03712 | 0.016024 |
| AN | Day 2 | Day 6B | 0.006667 | -0.02855 | 0.040509 |
| AN | Day 3 | Day 4 | -0.03891 | -0.09546 | 0.018231 |
| **AN *** | **Day 3** | **Day 5** | **-0.05862** | **-0.12382** | **-0.01568** |
| **AN *** | **Day 3** | **Day 6A** | **-0.05759** | **-0.11743** | **-0.01266** |
| AN | Day 3 | Day 6B | -0.04319 | -0.10796 | 0.004436 |
| AN | Day 4 | Day 5 | -0.01851 | -0.05499 | 0.007376 |
| AN | Day 4 | Day 6A | -0.01774 | -0.05426 | 0.009649 |
| AN | Day 4 | Day 6B | -0.00404 | -0.04572 | 0.032862 |
| AN | Day 5 | Day 6A | 0.00126 | -0.01714 | 0.02339 |
| AN | Day 5 | Day 6B | 0.014272 | -0.01005 | 0.047712 |
| AN | Day 6A | Day 6B | 0.012824 | -0.01104 | 0.047975 |
| N | Day 1 | Day 2 | 0.117715 | -0.03371 | 0.251403 |
| N | Day 1 | Day 3 | 0.032959 | -0.12282 | 0.176611 |
| N | Day 1 | Day 4 | 0.037213 | -0.11819 | 0.178162 |
| **N *** | **Day 1** | **Day 5** | **0.233974** | **0.093664** | **0.361136** |
| N | Day 1 | Day 6A | 0.074291 | -0.06796 | 0.222944 |
| **N *** | **Day 1** | **Day 6B** | **0.225396** | **0.078467** | **0.344971** |
| N | Day 2 | Day 3 | -0.0838 | -0.25582 | 0.063617 |
| N | Day 2 | Day 4 | -0.08126 | -0.23399 | 0.069743 |
| N | Day 2 | Day 5 | 0.113221 | -0.03656 | 0.24679 |
| N | Day 2 | Day 6A | -0.04351 | -0.19558 | 0.112185 |
| N | Day 2 | Day 6B | 0.104256 | -0.03075 | 0.249659 |
| N | Day 3 | Day 4 | 0.003656 | -0.16104 | 0.161508 |
| **N *** | **Day 3** | **Day 5** | **0.197768** | **0.051396** | **0.343886** |
| N | Day 3 | Day 6A | 0.03809 | -0.11595 | 0.21394 |
| **N *** | **Day 3** | **Day 6B** | **0.189104** | **0.048295** | **0.352102** |
| **N *** | **Day 4** | **Day 5** | **0.192473** | **0.039866** | **0.332024** |
| N | Day 4 | Day 6A | 0.038911 | -0.10945 | 0.209568 |
| **N *** | **Day 4** | **Day 6B** | **0.185669** | **0.041148** | **0.330322** |
| **N *** | **Day 5** | **Day 6A** | **-0.15711** | **-0.29838** | **-0.00992** |
| N | Day 5 | Day 6B | -0.01008 | -0.13812 | 0.111957 |
| **N *** | **Day 6A** | **Day 6B** | **0.148162** | **0.008968** | **0.293919** |

**Table S5.** Posterior summary of pairwise contrast comparisons for demeanor during the approachability test across days within the positive cue condition. Estimates represent the median posterior differences between conditions. Values in parentheses denote the 95% Highest Posterior Density (HPD) intervals. Pairwise comparisons marked with an asterisk (*) indicate statistically credible differences where the 95% HPD interval completely excludes zero.

| Demeanor | Day (A) | Day (B) | Median | Lower HPD | Upper HPD |
|---|---|---|---:|---:|---:|
| AF | Day 1 | Day 2 | -0.04508 | -0.16641 | 0.064963 |

| AF | Day 1 | Day 3 | -0.04862 | -0.16467 | 0.056448 |
|---|---|---|---|---|---|
| AF | Day 1 | Day 4 | 0.079222 | -0.0425 | 0.198735 |
| AF | Day 1 | Day 5 | 0.042295 | -0.07604 | 0.162162 |
| **AF *** | **Day 1** | **Day 6A** | **0.214969** | **0.090242** | **0.345903** |
| AF | Day 1 | Day 6B | 0.090731 | -0.03361 | 0.216692 |
| AF | Day 2 | Day 3 | -0.00344 | -0.13578 | 0.112722 |
| AF | Day 2 | Day 4 | 0.124492 | -0.01452 | 0.248873 |
| AF | Day 2 | Day 5 | 0.089726 | -0.04741 | 0.216967 |
| **AF *** | **Day 2** | **Day 6A** | **0.258195** | **0.116207** | **0.390042** |
| AF | Day 2 | Day 6B | 0.137458 | -0.00269 | 0.265558 |
| AF | Day 3 | Day 4 | 0.128228 | -0.00663 | 0.254951 |
| AF | Day 3 | Day 5 | 0.093137 | -0.03287 | 0.233277 |
| **AF *** | **Day 3** | **Day 6A** | **0.262154** | **0.122489** | **0.397364** |
| **AF *** | **Day 3** | **Day 6B** | **0.140274** | **0.010722** | **0.285938** |
| AF | Day 4 | Day 5 | -0.03786 | -0.16536 | 0.106608 |
| AF | Day 4 | Day 6A | 0.132617 | -0.01219 | 0.272946 |
| AF | Day 4 | Day 6B | 0.01149 | -0.13299 | 0.153189 |
| **AF *** | **Day 5** | **Day 6A** | **0.169895** | **0.027578** | **0.309644** |
| AF | Day 5 | Day 6B | 0.047521 | -0.09135 | 0.189449 |
| AF | Day 6A | Day 6B | -0.11996 | -0.26837 | 0.024438 |
| AG | Day 1 | Day 2 | -0.01154 | -0.04094 | 0.016208 |
| AG | Day 1 | Day 3 | -0.02048 | -0.04984 | 0.002396 |
| AG | Day 1 | Day 4 | -0.01547 | -0.04622 | 0.009321 |

| **AG *** | **Day 1** | **Day 5** | **-0.02497** | **-0.05736** | **-0.0036** |
|---|---|---|---|---|---|
| AG | Day 1 | Day 6A | -0.00952 | -0.03973 | 0.022305 |
| AG | Day 1 | Day 6B | -0.019 | -0.05021 | 0.006245 |
| AG | Day 2 | Day 3 | -0.00827 | -0.03908 | 0.014055 |
| AG | Day 2 | Day 4 | -0.00377 | -0.03208 | 0.027211 |
| AG | Day 2 | Day 5 | -0.0121 | -0.04488 | 0.005236 |
| AG | Day 2 | Day 6A | 0.00154 | -0.02701 | 0.037668 |
| AG | Day 2 | Day 6B | -0.00652 | -0.03657 | 0.019549 |
| AG | Day 3 | Day 4 | 0.004416 | -0.01669 | 0.030938 |
| AG | Day 3 | Day 5 | -0.00373 | -0.02537 | 0.013744 |
| AG | Day 3 | Day 6A | 0.010227 | -0.01254 | 0.042872 |
| AG | Day 3 | Day 6B | 0.001606 | -0.01998 | 0.025085 |
| AG | Day 4 | Day 5 | -0.00839 | -0.032 | 0.012236 |
| AG | Day 4 | Day 6A | 0.005592 | -0.02286 | 0.038179 |
| AG | Day 4 | Day 6B | -0.00319 | -0.02834 | 0.021235 |
| AG | Day 5 | Day 6A | 0.014497 | -0.00554 | 0.045412 |
| AG | Day 5 | Day 6B | 0.005228 | -0.01037 | 0.029953 |
| AG | Day 6A | Day 6B | -0.0088 | -0.04051 | 0.014459 |
| AN | Day 1 | Day 2 | 0.000523 | -0.00635 | 0.011163 |
| AN | Day 1 | Day 3 | 0.002997 | -0.00549 | 0.015739 |
| AN | Day 1 | Day 4 | 0.00049 | -0.00815 | 0.010566 |
| AN | Day 1 | Day 5 | -0.00084 | -0.00743 | 0.007505 |
| AN | Day 1 | Day 6A | -0.00024 | -0.00751 | 0.010178 |
| AN | Day 1 | Day 6B | 0.004005 | -0.00618 | 0.018513 |
| AN | Day 2 | Day 3 | 0.002314 | -0.01232 | 0.016643 |
| AN | Day 2 | Day 4 | -2.05E-05 | -0.01333 | 0.011447 |
| AN | Day 2 | Day 5 | -0.0014 | -0.0136 | 0.009263 |
| AN | Day 2 | Day 6A | -0.00076 | -0.0137 | 0.010814 |
| AN | Day 2 | Day 6B | 0.00325 | -0.0114 | 0.018755 |
| AN | Day 3 | Day 4 | -0.00239 | -0.0198 | 0.009772 |
| AN | Day 3 | Day 5 | -0.00395 | -0.01933 | 0.007515 |
| AN | Day 3 | Day 6A | -0.00298 | -0.01999 | 0.008565 |
| AN | Day 3 | Day 6B | 0.000754 | -0.0149 | 0.019005 |
| AN | Day 4 | Day 5 | -0.00148 | -0.01427 | 0.008726 |
| AN | Day 4 | Day 6A | -0.00085 | -0.01412 | 0.011179 |
| AN | Day 4 | Day 6B | 0.003101 | -0.01223 | 0.020371 |
| AN | Day 5 | Day 6A | 0.000627 | -0.00805 | 0.01433 |
| AN | Day 5 | Day 6B | 0.00469 | -0.00791 | 0.021965 |
| AN | Day 6A | Day 6B | 0.00388 | -0.00818 | 0.023436 |
| N | Day 1 | Day 2 | 0.057647 | -0.06896 | 0.178078 |
| N | Day 1 | Day 3 | 0.067678 | -0.04749 | 0.189521 |
| N | Day 1 | Day 4 | -0.06474 | -0.18363 | 0.067275 |
| N | Day 1 | Day 5 | -0.01636 | -0.14267 | 0.105925 |
| **N *** | **Day 1** | **Day 6A** | **-0.20458** | **-0.34206** | **-0.07446** |
| N | Day 1 | Day 6B | -0.07596 | -0.2174 | 0.051485 |
| N | Day 2 | Day 3 | 0.011458 | -0.12994 | 0.139623 |
| N | Day 2 | Day 4 | -0.12015 | -0.25288 | 0.027227 |
| N | Day 2 | Day 5 | -0.07289 | -0.21081 | 0.064903 |
| **N *** | **Day 2** | **Day 6A** | **-0.26114** | **-0.39919** | **-0.11303** |
| N | Day 2 | Day 6B | -0.1324 | -0.2672 | 0.01367 |
| N | Day 3 | Day 4 | -0.13262 | -0.26202 | 0.014509 |
| N | Day 3 | Day 5 | -0.0827 | -0.21598 | 0.063426 |
| **N *** | **Day 3** | **Day 6A** | **-0.26872** | **-0.40717** | **-0.12392** |
| **N *** | **Day 3** | **Day 6B** | **-0.14297** | **-0.29559** | **-0.00787** |
| N | Day 4 | Day 5 | 0.049064 | -0.09447 | 0.189439 |
| N | Day 4 | Day 6A | -0.13674 | -0.28477 | 0.008105 |
| N | Day 4 | Day 6B | -0.01156 | -0.17112 | 0.120806 |
| **N *** | **Day 5** | **Day 6A** | **-0.18875** | **-0.32785** | **-0.0399** |
| N | Day 5 | Day 6B | -0.06002 | -0.20857 | 0.082267 |
| N | Day 6A | Day 6B | 0.126067 | -0.02044 | 0.28109 |

**Table S6.** Posterior summary of pairwise contrast comparisons for demeanor during the approachability test across days between male and female dogs. Estimates represent the median posterior differences between conditions. Values in parentheses denote the 95% Highest Posterior Density (HPD) intervals. Pairwise comparisons marked with an asterisk (*) indicate statistically credible differences where the 95% HPD interval completely excludes zero.

| Day | Demeanor | Median | Lower HPD | Upper HPD |
|---|---|---|---|---|
| **Day 1** | **AF *** | **-0.11303** | **-0.20295** | **-0.02401** |
| Day 1 | AG | -0.02262 | -0.06208 | 0.006543 |
| Day 1 | AN | -0.00116 | -0.01125 | 0.007855 |
| **Day 1** | **N *** | **0.139329** | **0.038841** | **0.238688** |
| **Day 2** | **AF *** | **-0.10029** | **-0.1888** | **-0.02558** |
| Day 2 | AG | -0.01702 | -0.05219 | 0.004266 |
| Day 2 | AN | -0.00131 | -0.01288 | 0.009723 |
| **Day 2** | **N *** | **0.121771** | **0.036145** | **0.221541** |
| **Day 3** | **AF *** | **-0.10899** | **-0.19651** | **-0.03063** |
| Day 3 | AG | -0.00848 | -0.02586 | 0.00438 |
| Day 3 | AN | -0.00517 | -0.03543 | 0.026554 |
| **Day 3** | **N *** | **0.124252** | **0.028257** | **0.217103** |
| **Day 4** | **AF *** | **-0.12154** | **-0.21372** | **-0.02565** |
| Day 4 | AG | -0.00753 | -0.02505 | 0.004037 |
| Day 4 | AN | -0.00162 | -0.01702 | 0.013559 |
| **Day 4** | **N *** | **0.133394** | **0.031179** | **0.231554** |
| **Day 5** | **AF *** | **-0.10708** | **-0.19275** | **-0.02531** |
| Day 5 | AG | -0.00419 | -0.0159 | 0.002424 |
| Day 5 | AN | -0.00065 | -0.00771 | 0.004654 |
| **Day 5** | **N *** | **0.113404** | **0.025622** | **0.201423** |
| **Day 6A** | **AF *** | **-0.12542** | **-0.21518** | **-0.02365** |
| Day 6A | AG | -0.01132 | -0.03459 | 0.009093 |
| Day 6A | AN | -0.00029 | -0.00781 | 0.007419 |
| **Day 6A** | **N *** | **0.13938** | **0.043409** | **0.244051** |
| **Day 6B** | **AF *** | **-0.10223** | **-0.18238** | **-0.01907** |

| Day 6B | AG | -0.00531 | -0.01744 | 0.003456 |
|---|---|---|---|---|
| Day 6B | AN | -0.00218 | -0.01806 | 0.013079 |
| **Day 6B** | **N *** | **0.111326** | **0.02685** | **0.207613** |

**Table S7.** Posterior summary of pairwise contrast comparisons for demeanor during the test phase between cue conditions. Estimates represent the median posterior differences between conditions. Values in parentheses denote the 95% Highest Posterior Density (HPD) intervals. Pairwise comparisons marked with an asterisk (*) indicate statistically credible differences where the 95% HPD interval completely excludes zero.

| Day | Demeanor | Median | Lower HPD | Upper HPD |
|---|---|---|---|---|
| Day 1 | AF | -0.01959 | -0.12918 | 0.0922 |
| **Day 1** | **AG *** | **0.04463** | **0.00307** | **0.096146** |
| Day 1 | AN | 0.018786 | -0.00075 | 0.051418 |
| Day 1 | N | -0.04931 | -0.18075 | 0.086499 |
| Day 2 | AF | -0.06104 | -0.16137 | 0.040124 |
| **Day 2** | **AG *** | **0.070572** | **0.00679** | **0.151067** |
| Day 2 | AN | 0.019591 | -0.00107 | 0.06119 |
| Day 2 | N | -0.03464 | -0.17339 | 0.102054 |
| Day 3 | AF | -0.0373 | -0.16927 | 0.075703 |
| Day 3 | AG | 0.022928 | -0.0023 | 0.064895 |
| Day 3 | AN | 0.028587 | -0.00108 | 0.08019 |

| Day 3 | N | -0.01809 | -0.16576 | 0.113479 |
|---|---|---|---|---|
| Day 4 | AF | -0.06744 | -0.20829 | 0.051621 |
| Day 4 | AG | 0.005602 | -0.01419 | 0.033843 |
| Day 4 | AN | 0.013791 | -0.00513 | 0.044727 |
| Day 4 | N | 0.044085 | -0.08596 | 0.198741 |
| Day 5 | AF | -0.07972 | -0.17793 | 0.039496 |
| Day 5 | AG | 0.014477 | -0.00532 | 0.046131 |
| Day 5 | AN | 0.007722 | -0.00613 | 0.033106 |
| Day 5 | N | 0.051498 | -0.07192 | 0.16885 |

**Table S8.** Posterior summary of pairwise contrast comparisons for demeanor during the test phase across days within the threatening cue condition. Estimates represent the median posterior differences between conditions. Values in parentheses denote the 95% Highest Posterior Density (HPD) intervals. Pairwise comparisons marked with an asterisk (*) indicate statistically credible differences where the 95% HPD interval completely excludes zero.

| Demeanor | Day (A) | Day (B) | Median | Lower HPD | Upper HPD |
|---|---|---|---|---|---|
| **AF *** | **Day 1** | **Day 2** | **-0.1094** | **-0.21288** | **-0.02149** |
| AF | Day 1 | Day 3 | -0.0301 | -0.12575 | 0.083157 |
| AF | Day 1 | Day 4 | -0.03809 | -0.14127 | 0.071349 |
| **AF *** | **Day 1** | **Day 5** | **-0.09889** | **-0.2011** | **-0.00527** |
| AF | Day 2 | Day 3 | 0.081926 | -0.02045 | 0.18625 |
| AF | Day 2 | Day 4 | 0.071566 | -0.02253 | 0.187118 |
| AF | Day 2 | Day 5 | 0.012514 | -0.08159 | 0.101347 |
| AF | Day 3 | Day 4 | -0.00752 | -0.12467 | 0.108323 |
| AF | Day 3 | Day 5 | -0.06894 | -0.17051 | 0.043316 |
| AF | Day 4 | Day 5 | -0.05859 | -0.16546 | 0.043838 |
| AG | Day 1 | Day 2 | 0.029086 | -0.0326 | 0.108684 |
| AG | Day 1 | Day 3 | -0.02863 | -0.08697 | 0.012127 |
| **AG *** | **Day 1** | **Day 4** | **-0.04291** | **-0.09708** | **-0.00344** |
| AG | Day 1 | Day 5 | -0.03761 | -0.08953 | 0.004076 |
| **AG *** | **Day 2** | **Day 3** | **-0.06084** | **-0.13618** | **-0.00431** |
| **AG *** | **Day 2** | **Day 4** | **-0.07482** | **-0.14755** | **-0.01333** |
| **AG *** | **Day 2** | **Day 5** | **-0.06972** | **-0.14458** | **-0.01183** |
| AG | Day 3 | Day 4 | -0.01204 | -0.05498 | 0.018505 |
| AG | Day 3 | Day 5 | -0.00763 | -0.04724 | 0.03169 |
| AG | Day 4 | Day 5 | 0.004483 | -0.02481 | 0.038795 |
| AN | Day 1 | Day 2 | -0.00068 | -0.03553 | 0.032024 |
| AN | Day 1 | Day 3 | 0.009114 | -0.02622 | 0.055154 |
| AN | Day 1 | Day 4 | -0.00627 | -0.03799 | 0.022994 |
| AN | Day 1 | Day 5 | -0.01278 | -0.0466 | 0.011558 |
| AN | Day 2 | Day 3 | 0.009708 | -0.0303 | 0.0573 |
| AN | Day 2 | Day 4 | -0.00528 | -0.03872 | 0.029101 |
| AN | Day 2 | Day 5 | -0.01184 | -0.04854 | 0.020036 |
| AN | Day 3 | Day 4 | -0.01538 | -0.06336 | 0.018987 |
| AN | Day 3 | Day 5 | -0.02208 | -0.07315 | 0.005254 |
| AN | Day 4 | Day 5 | -0.00593 | -0.03914 | 0.017181 |
| N | Day 1 | Day 2 | 0.078125 | -0.04174 | 0.200484 |
| N | Day 1 | Day 3 | 0.050776 | -0.08456 | 0.164849 |
| N | Day 1 | Day 4 | 0.091861 | -0.03147 | 0.215125 |
| **N *** | **Day 1** | **Day 5** | **0.152884** | **0.03962** | **0.268995** |
| N | Day 2 | Day 3 | -0.02981 | -0.151 | 0.110388 |
| N | Day 2 | Day 4 | 0.013216 | -0.10304 | 0.148268 |
| N | Day 2 | Day 5 | 0.073578 | -0.04116 | 0.196462 |
| N | Day 3 | Day 4 | 0.039697 | -0.08661 | 0.16807 |
| N | Day 3 | Day 5 | 0.102708 | -0.01563 | 0.229728 |
| N | Day 4 | Day 5 | 0.061043 | -0.05791 | 0.175934 |

**Table S9.** Posterior summary of pairwise contrast comparisons for demeanor during the test phase across days within the positive cue condition. Estimates represent the median posterior differences between conditions. Values in parentheses denote the 95% Highest Posterior Density (HPD) intervals. Pairwise comparisons marked with an asterisk (*) indicate statistically credible differences where the 95% HPD interval completely excludes zero.

| Demeanor | Day (A) | Day (B) | Median | Lower HPD | Upper HPD |
|---|---|---|---|---|---|
| AF | Day 1 | Day 2 | -0.07003 | -0.16314 | 0.028794 |
| AF | Day 1 | Day 3 | -0.01272 | -0.11465 | 0.083964 |
| AF | Day 1 | Day 4 | 0.012069 | -0.09068 | 0.118846 |
| AF | Day 1 | Day 5 | -0.04015 | -0.13535 | 0.062917 |
| AF | Day 2 | Day 3 | 0.056863 | -0.04539 | 0.1665 |
| AF | Day 2 | Day 4 | 0.079959 | -0.02824 | 0.188675 |
| AF | Day 2 | Day 5 | 0.029407 | -0.07746 | 0.135764 |
| AF | Day 3 | Day 4 | 0.02253 | -0.08457 | 0.149977 |
| AF | Day 3 | Day 5 | -0.02783 | -0.13941 | 0.073424 |
| AF | Day 4 | Day 5 | -0.05097 | -0.16626 | 0.052298 |
| AG | Day 1 | Day 2 | 0.005635 | -0.01322 | 0.03394 |
| AG | Day 1 | Day 3 | -0.00818 | -0.02686 | 0.002686 |
| AG | Day 1 | Day 4 | -0.00519 | -0.02074 | 0.012914 |
| AG | Day 1 | Day 5 | -0.00836 | -0.02611 | 0.004195 |
| AG | Day 2 | Day 3 | -0.01477 | -0.04517 | 0.002141 |
| AG | Day 2 | Day 4 | -0.01129 | -0.04155 | 0.00732 |
| AG | Day 2 | Day 5 | -0.01469 | -0.04237 | 0.002685 |
| AG | Day 3 | Day 4 | 0.002896 | -0.01011 | 0.019221 |
| AG | Day 3 | Day 5 | -0.00032 | -0.01479 | 0.011586 |
| AG | Day 4 | Day 5 | -0.00323 | -0.01991 | 0.009716 |
| AN | Day 1 | Day 2 | -0.00168 | -0.01101 | 0.006803 |
| AN | Day 1 | Day 3 | 9.93E-05 | -0.0101 | 0.011594 |
| AN | Day 1 | Day 4 | -0.00172 | -0.01067 | 0.008254 |
| AN | Day 1 | Day 5 | -0.00246 | -0.01298 | 0.005287 |
| AN | Day 2 | Day 3 | 0.001883 | -0.01005 | 0.014088 |
| AN | Day 2 | Day 4 | -1.91E-05 | -0.0108 | 0.010878 |
| AN | Day 2 | Day 5 | -0.00068 | -0.01074 | 0.008942 |
| AN | Day 3 | Day 4 | -0.00178 | -0.01469 | 0.00976 |
| AN | Day 3 | Day 5 | -0.00264 | -0.01549 | 0.006935 |
| AN | Day 4 | Day 5 | -0.0007 | -0.0117 | 0.008162 |
| N | Day 1 | Day 2 | 0.064413 | -0.05233 | 0.157674 |
| N | Day 1 | Day 3 | 0.021138 | -0.08051 | 0.125809 |
| N | Day 1 | Day 4 | -0.00406 | -0.11891 | 0.103622 |
| N | Day 1 | Day 5 | 0.052267 | -0.05336 | 0.159337 |
| N | Day 2 | Day 3 | -0.04288 | -0.1582 | 0.066968 |
| N | Day 2 | Day 4 | -0.06791 | -0.1876 | 0.038857 |
| N | Day 2 | Day 5 | -0.01185 | -0.12597 | 0.101841 |
| N | Day 3 | Day 4 | -0.02418 | -0.15387 | 0.086733 |
| N | Day 3 | Day 5 | 0.031527 | -0.08582 | 0.134296 |
| N | Day 4 | Day 5 | 0.055899 | -0.04477 | 0.181303 |

**Table S10.** Posterior summary of pairwise contrast comparisons for demeanor during test phase across days between male and female dogs. Estimates represent the median posterior differences between conditions. Values in parentheses denote the 95% Highest Posterior Density (HPD) intervals. Pairwise comparisons marked with an asterisk (*) indicate statistically credible differences where the 95% HPD interval completely excludes zero.

| Day | Demeanor | Median | Lower HPD | Upper HPD |
|---|---|---:|---:|---:|
| Day 1 | AF | -0.02513 | -0.09037 | 0.041865 |
| Day 1 | AG | -0.00336 | -0.03051 | 0.025078 |
| Day 1 | AN | -0.00221 | -0.01969 | 0.011509 |
| Day 1 | N | 0.030314 | -0.05174 | 0.099683 |
| Day 2 | AF | -0.01714 | -0.06822 | 0.026672 |
| Day 2 | AG | -0.00186 | -0.04587 | 0.034272 |
| Day 2 | AN | -0.00141 | -0.01617 | 0.013644 |
| Day 2 | N | 0.02073 | -0.04164 | 0.088924 |
| Day 3 | AF | -0.02287 | -0.0866 | 0.038729 |
| Day 3 | AG | -0.00114 | -0.01612 | 0.013224 |
| Day 3 | AN | -0.00249 | -0.02391 | 0.015609 |
| Day 3 | N | 0.02683 | -0.04198 | 0.097686 |
| Day 4 | AF | -0.02483 | -0.08429 | 0.042459 |
| Day 4 | AG | -0.00022 | -0.01101 | 0.011565 |
| Day 4 | AN | -0.00098 | -0.01254 | 0.011622 |
| Day 4 | N | 0.025797 | -0.04573 | 0.089737 |
| Day 5 | AF | -0.02066 | -0.07472 | 0.033336 |
| Day 5 | AG | -0.00018 | -0.01166 | 0.010988 |
| Day 5 | AN | -0.0005 | -0.00991 | 0.006656 |
| Day 5 | N | 0.021773 | -0.03715 | 0.079731 |